\documentclass[10pt]{iopart}
\pdfoutput=1
\usepackage[english]{babel}

\usepackage{amssymb}
\usepackage{bm}
\usepackage{graphics}
\usepackage{graphicx}
\usepackage{iopams} 
\usepackage{float}

\usepackage[utf8]{inputenc}
\usepackage[T1]{fontenc}
\usepackage{ae,aecompl}

\usepackage{bm} 

\usepackage{stackrel}

\usepackage[usenames,dvipsnames]{xcolor}

\usepackage{color}
\definecolor{darkgreen}{rgb}{0,0.6,0}
\definecolor{darkblue}{rgb}{0,0,0.6}
\definecolor{darkred}{rgb}{0.6,0,0}
\definecolor{darkpurple}{rgb}{0.5,0,0.5}

\usepackage{hyperref}

\hypersetup{
bookmarksopen=true,
pdftitle=Effective dynamics for large deviations of Langevin processes in the weak-noise limit,
pdfauthor=Nicolás Tizón-Escamilla and Vivien Lecomte and Eric Bertin, 
pdftoolbar=false, 
pdfstartview={FitH},		
pdfmenubar=true,			
pdfhighlight=/O,			
colorlinks=true,			
urlcolor=darkblue,
citecolor=darkblue,		
linkcolor=darkpurple	
}

\usepackage[left=1.0in,right=1.0in,top=1.0in,bottom=1.0in,includeheadfoot]{geometry}

\newcommand{\tf}{{t_{\rm{f}}}}

\newcommand{\WW}{\mathbb W}
\newcommand{\Sla}{S_{\mkern-1.5mu \lambda}}

\newcommand{\ee}{{\rm{e}}}
\newcommand{\cc}{{\rm{c}}}
\newcommand{\ii}{{\rm{i}}}

\newcommand{\xstar}{x^{\star\mkern-1.5mu}}
\renewcommand{\epsilon}{\varepsilon}

\providecommand{\tfrac}[2]{\textnormal{$\frac{#1}{#2}$}}
\providecommand{\eqref}[1]{\eref{#1}}
\providecommand{\text}[1]{{\rm{#1}}}
\providecommand{\operatorname}[1]{{\,\rm{#1}}}

\newcommand{\eff}{{\mkern+1.mu\text{eff}\mkern-1.5mu}}

\providecommand{\mathds}[1]{\mathbf{#1}}

\newenvironment{align}
 {\begin{eqnarray}}
 {\end{eqnarray}\ignorespacesafterend}

\begin{document}

\vspace*{-2cm}

\title{%
  Effective driven dynamics for one-dimensional conditioned Langevin processes in the weak-noise limit
}


\author{Nicol\'as Tiz\'on-Escamilla$^{1,2}$, Vivien Lecomte$^2$ and Eric Bertin$^2$}

\address{$^1$
Departamento de Electromagnetismo y F\'{\i}sica de la Materia,
and Instituto Carlos I de F\'{\i}sica Te\'orica y Computacional. Universidad de Granada. E-18071 Granada. Spain
}
\address{$^2$
Univ.~Grenoble Alpes, CNRS, LIPhy, F-38000 Grenoble, France
}
 


\begin{abstract}
In this work we focus on fluctuations of time-integrated observables for a particle diffusing in a one-dimensional periodic potential in the weak-noise asymptotics.
Our interest goes to rare trajectories presenting an atypical value of the observable, that we study through a biased dynamics in a large-deviation framework.
We determine explicitly the effective probability-conserving dynamics which makes rare trajectories of the original dynamics become typical trajectories of the effective one.
Our approach makes use of a weak-noise path-integral description in which the action is minimised by the rare trajectories of interest.
For `current-type' additive observables, we find the emergence of a propagative trajectory minimising the action for large enough deviations, revealing the existence of a dynamical phase transition at a fluctuating level. In addition, we provide a new method to determine the scaled cumulant generating function of the observable without having to optimise the action. It allows one to show that the weak-noise and the large-time limits commute in this problem.
Finally, we show how the biased dynamics can be mapped in practice to an effective driven dynamics, which takes the form of a driven Langevin dynamics in an effective potential. The non-trivial shape of this effective potential is key to understand the link between the dynamical phase transition in the large deviations of current and the standard depinning transition of a particle in a tilted potential.
%
%

\vspace{2pc}

\end{abstract}


\begin{small}
	\tableofcontents
\end{small}

\newpage
\section{Introduction}
\label{sec:introduction}

Traditional approaches in statistical physics are based on the study of the probability distribution of microscopic configurations at a given time~\cite{landaustat}.
Although such approaches have been very successful at equilibrium where configurations with the same energy are distributed uniformly in an isolated system, one is faced with difficulties when considering the statistics of configurations in nonequilibrium steady-states, as this statistics is in general non-uniform and unknown.
It has been realised in the last decades that a more general space-time formulation, which deals with the statistics of full trajectories (that is, configurations as a function of time on a large time window) could be formulated in a quite general way, even for non-equilibrium systems~\cite{onsager53a,onsager53b}.
Moreover, the large-deviation formalism provides an efficient framework to formulate the problem~\cite{ellis_entropy_1985,ellis_overview_1995,ellis_theory_1999,bertini15a,touchette_large_2009}. The large-deviation formalism is particularly useful for instance to evaluate the statistics of time-integrated observables (\emph{e.g.}, particle current or dynamical activity), which are natural observables when characterising the statistics of trajectories~\cite{bertini15a,touchette_large_2009,derrida07a,bertini05a,bertini06a,hurtado14a,derrida_exact_1998,lebowitz_gallavotticohen-type_1999,derrida_universal_1999}.
One can for instance consider a modified equilibrium statistics of trajectories conditioned to a given value of a time-integrated observable, like the average particle current.
It is then of interest to ask whether this artificial, biased dynamics shares some similarities with (or even could be mapped to) a `real' non-equilibrium dynamics. In other words, does a physical force which drives a system into a non-equilibrium state (and thus generates a given current) selects all trajectories having a given average current in a uniform way?

In practice, fixing a given average value of an integrated observable is done by introducing a conjugated Lagrange multiplier, in the same way as, at equilibrium, temperature fixes the average energy in the canonical ensemble~\cite{bertini15a}.
This Lagrange multiplier enters the definition of a `deformed' Markov operator that describes the biased dynamics.
A well-know difficulty is that this deformed Markov operator no longer conserves probability, and cannot straightforwardly be interpreted as describing a \emph{bona fide} probability-preserving dynamics.
It has however been shown~\cite{miller_convexity_1961,simon_construction_2009,popkov_asep_2010,jack_large_2010} how a relatively simple but formal transformation of the deformed Markov operator allows one to define a closely related probability-conserving Markov operator.

The goal of this paper is to go beyond formal definitions and to show how such a transformation can be used in practice, on the example of a particle diffusing in a potential in a one-dimension. Weak-noise, Wentzel--Kramers--Brillouin (WKB) \cite{Olver1974} type approximations allow us to obtain an explicit mapping onto a non-equilibrium dynamics. 
We extend a recent work in which the large deviations of the current were studied in the weak-noise asymptotics~\cite{tsobgni_nyawo_large_2016} to the case of generic time-integrated observables%
\footnote{%
During the preparation of this manuscript, we became aware that works parallel to ours were completed using similar approaches~\cite{derrida_sadhu_2018,derrida_proesmans_2018}.
}.
Besides, we find explicit generic solutions to the optimisation principle that governs the value of large deviation functions.

The appearance of singularities (as non-differentiabilities) is known to occur in the quasi-potential of non-equilibrium dynamics in the weak-noise limit~\cite{graham_weak-noise_1985,jauslin_nondifferentiable_1987,dykman_observable_1994}, when considering the large-deviation scaling of the steady-state distribution.
In the context of time-integrated observables, the occurrence of another type of singularities has been reported during the last decade in varied systems~\cite{bodineau05a,bertini05a,bodineau2007cumulants,lecomte_thermodynamic_2007,PhysRevLett.98.195702,hurtado14a,nyawo_minimal_2016}, and corresponds to the type of dynamical phase transitions we are interested in in this paper. These describe how the trajectories that lead to an atypical value of the time-integrated observable can change from one class to another when varying the value of this observable.

The paper is organised as follows.
In Sec.~\ref{sec:rare-traj-cond}, we define the Langevin dynamics and reformulate it in a path-integral framework, to be able to bias the dynamics by a given value of the integrated additive observable considered. We also introduce the large deviation form of the action at large time, leading to the definition of the scaled cumulant generating function.
In Sec.~\ref{sec:case-peri-syst}, we show how the scaled cumulant generating function can be evaluated explicitly, in the small noise limit, using a saddle-node calculation.
Finally, in Sec.~\ref{sec:effective-dyn}, we use the knowledge of the scaled cumulant generating function to derive an effective physically driven dynamics that leads to the same statistics of trajectories, and discuss its interpretation.

\section{Rare trajectories: conditioning or biasing the dynamics}
\label{sec:rare-traj-cond}

We present in this section the Langevin dynamics we focus on, and the type of the observables whose distribution we are interested in.
We refer the reader to existing reviews~\cite{touchette_large_2009,touchette_introduction_2017} for generalisations for instance to mixed Langevin and Markov jump processes.

\subsection{Langevin dynamics and additive observables}
\label{sec:lang-dynam-addit}

Consider a particle of position $x(t)$ at time $t$ subjected to a force $F(x)$ and a thermal noise $\eta(t)$.
In the overdamped limit, the evolution of its position is described by the Langevin equation 
\begin{equation}
\label{eq:Langevinepsilon}
 \dot{x}(t)=F(x(t))+\sqrt{\epsilon}\,\eta(t)\,\,,
\end{equation}
where $\eta(t)$ is a Gaussian white noise of average $\langle\eta(t)\rangle=0$ and correlation function $\langle\eta(t)\eta(t')\rangle=\delta(t-t')$. 
Our interest goes to observables depending on the trajectory on a time window $[0,\tf]$ and taking the form\footnote{%
  The stochastic integral in~\eqref{eq:obs} is taken in the Stratonovich convention.
}%
\begin{equation}
 \label{eq:obs}
 A(\tf)=\int_0^{\tf} h(x(t))\,dt+\int_0^{\tf} g(x(t))\,\dot{x}(t)\,dt\,\,.
\end{equation}
Examples of such an \emph{additive} observable encompass time-integrated current, work, entropy production, or activity for specific choices of the functions $h(x)$ and $g(x)$ (see~\cite{touchette_introduction_2017} for examples).
One is interested in the distribution $P(A,\tf)$ of the observable $A$ at time~$\tf$, in the weak-noise and/or the large-time limit.
In these asymptotics, the scaling form of $P(A,\tf)$ is described within the framework of the large-deviation theory, which has witnessed a tremendous development in the last decades both in mathematics (within the Donsker--Varadhan~\cite{donsker_asymptotic_1975,donsker_asymptotic_1975-1,donsker_asymptotic_1976,donsker_asymptotic_1983}, the Gärtner--Ellis~\cite{gartner_large_1977,ellis_large_1984} and the Freidlin--Wentzell approaches~\cite{freidlin_random_2012}) and in statistical physics in the study of many example systems (see for instance~\cite{derrida_exact_1998,kurchan_fluctuation_1998,bodineau_current_2004,bertini_current_2005,lecomte_chaotic_2005,lecomte_thermodynamic_2007}).
While we refer the reader to many existing reviews for a general presentation of this approach~\cite{ellis_entropy_1985,ellis_overview_1995,ellis_theory_1999,touchette_large_2009,bertini_macroscopic_2015,touchette_introduction_2017},
we provide here for completeness a self-contained presentation of the tools used in our context.

The Langevin equation~\eqref{eq:Langevinepsilon} is equivalently described by the Onsager--Machlup weight of a trajectory $[x(t)]_{0\leq t\leq\tf}$ of duration $\tf$
\begin{equation}
 \mathcal{P}[x(t),\tf]\propto \ee^{-\frac{1}{2\epsilon}\int_0^{\tf} (\dot{x}-F)^2\,dt}
\end{equation}
(valid in the weak-noise asymptotics $\epsilon\to 0$, since we are working with the Stratonovich convention)
or by the Fokker--Planck equation for the evolution of the probability $P(x,t)$ of finding the particle at a position $x$ at time $t$, which takes the form
\begin{equation}
 \partial_t{P}(x,t)=\WW P(x,t)\,.
\end{equation}
The Fokker--Planck operator $\WW$ reads:
\begin{equation}\label{eq:FPope}
 \WW \cdot=-\partial_x(F(x)\cdot)+\frac{1}{2}\epsilon\partial^2_x\cdot\,\,.
\end{equation}
Note that the conservation of probability reads $\langle -|\WW=0$ where $\langle -|$ is the flat vector with all components equal to 1 (\emph{i.e.}~$\langle -|x\rangle=1$ for all $x$)%
\footnote{%
Here we use a bra-ket notation to describe the vector space on which operators such as $\WW$ act, with $|x\rangle$ the state representing the particle at position $x$ and  $\langle x|$ its transpose. These define the canonical scalar product $\langle x|x'\rangle=\delta(x-x')$.
}%
.
In other words, $|-\rangle$ is a left eigenvector of $\WW$ of eigenvalue $0$. 
Our aim is to use and extend these descriptions in order to understand the distribution of the observable $A$.

\subsection{Path-integral and Fokker--Planck representations}
\label{sec:path-integral-fokker}

We aim at characterising the physical features of trajectories $[x(t)]_{0\leq t\leq\tf}$ presenting an arbitrary (for instance, atypical) value of the observable, $A$.
One way to proceed is to determine the probability $P(x,A,\tf)$ of the particle to be in position $x$ at time $t_f$, while having observed a value $A$ of the additive observable~(\ref{eq:obs}) on the time window $[0,\tf]$. It reads as follows
\begin{equation}
 \label{eq:defPAxtf}
 {P}(x,A,\tf) = 
\left< \int^{x(\tf)=x}_{x(0)}\mathcal{D}x\:\delta({A}-A(\tf))\:\ee^{-\frac{1}{2\epsilon}\int_0^{\tf} (\dot{x}-F)^2\,dt} \right>_{\!\!\! x(0)}
\end{equation}
where the notation $\langle \dots \rangle_{x(0)}$ indicates an average over the initial position $x(0)$ with a distribution $P_\ii(x)$. We implicitly assume that this is the case in the following, except otherwise indicated.

It is difficult in general to determine ${P}(x,A,\tf)$ or even to write a closed equation for this `microcanonical' probability. 
Following Varadhan~\cite{varadhan_asymptotic_1966}, one performs a Laplace transform and introduces the \emph{biased} distribution
\begin{align}
\!\!\!\!\!\!\!\!\!\!\!\!\!\!\!\! \hat{{P}}(x,\lambda,\tf)
  &=
 \int  \ee^{-\frac{\lambda}{\epsilon}{A}}\:   {P}(x,A,\tf) \:dA
\label{eq:Phatlambdadef}
\\
  &= \left< \int^{x(\tf)=x}_{x(0)}\mathcal{D}x \:\ee^{-\frac{1}{\epsilon}\big[\lambda A(\tf)+\frac 12 \int_0^{\tf} (\dot{x}-F)^2\,dt\big]} \right>_{\!\!\! x(0)}
\equiv
  \left< \int^{x(\tf)=x}_{x(0)} \mathcal{D}x \: \ee^{-\frac{1}{\epsilon}\Sla[x(t),\tf]} \right>_{\!\!\! x(0)}
\label{eq:Phatlambdapathintegral}
\end{align}
where $\Sla[x(t),\tf]$ is defined as
\begin{equation}
\Sla[x(t),\tf] = \lambda A(\tf)+\frac 12 \int_0^{\tf} (\dot{x}-F)^2\,dt
= \int_0^{\tf} \bigg\lbrace{ \frac 12 (\dot{x}-F)^2}+ \lambda \big( h + \dot x g\big)\bigg\rbrace\;dt
\label{eq:S-lambda1}
\end{equation}
with $F \equiv F\big(x(t)\big)$, $g \equiv g\big(x(t)\big)$ and $h \equiv h\big(x(t)\big)$ to lighten notations.
This defines a `canonical' version of the problem, where trajectories are biased by an exponential factor $\ee^{-\frac{\lambda}{\epsilon}{A(\tf)}}$ on the time window $[0,\tf]$.
In the large $\tf$ limit, as detailed below, the conditioned distribution corresponding to~\eqref{eq:defPAxtf} and the biased distribution~\eqref{eq:Phatlambdadef} become asymptotically equivalent (as studied in great depth by Chétrite and Touchette~\cite{chetrite_nonequilibrium_2013,chetrite_nonequilibrium_2015}), provided that the value of $\lambda$ is well chosen as a function of $A$, as in any change of ensemble\footnote{%
This implies some requirement on the convexity of a large deviation function, as we explain below.
}.
It is known that the evolution in time of ${\hat P}(x,\lambda,\tf)$ reads $\partial_t\hat P=\WW_\lambda\hat P$ with
a biased Fokker--Planck operator
\begin{equation}
  \WW _\lambda\cdot=-\partial_x\big((F-\lambda g)\,\cdot\big)+\frac{1}{2}\epsilon\partial^2_x\cdot+\frac{\lambda}{\epsilon}\bigg(\frac{\lambda}{2}g^2-gF-h \bigg)\mathds{1}\,.
\label{eq:Wlambda}
\end{equation}
This result can be derived by using a generalised Feynman--Kac formula~\cite{chetrite_nonequilibrium_2013,chetrite_nonequilibrium_2015}, or by remarking that the action $\Sla[x(t),\tf]$ in~\eqref{eq:S-lambda1} is equivalently written as:
\begin{align}
 \Sla[x(t),\tf]
=\int_0^{\tf} \bigg\lbrace{\frac 12 (\dot{x}-F+\lambda g)^2}-\lambda\Big(\frac{\lambda }{2} g^2-gF-h \Big)\bigg\rbrace\;dt\;.
\label{eq:S-lambda2}
\end{align}
This rewriting~(\ref{eq:S-lambda2}) of the action~(\ref{eq:S-lambda1}) is the only factorisation of the $\lambda\dot x g$ contribution into the square term of the action. Using the classical Feynman--Kac formula, the form of the biased operator~\eqref{eq:Wlambda} is then directly read from~(\ref{eq:S-lambda2}), since this action describes the trajectorial probability of a process $x(t)$ obeying a Langevin equation
\begin{equation} \label{eq:Langevin:modified}
\dot{x} = F(x)-\lambda g(x)+\sqrt{\epsilon}\,\eta
\end{equation}
 which is biased by a trajectorial weight
\begin{equation}
  \exp\Big\{
  \int_0^{\tf} 
  \frac{\lambda}{\epsilon}
  \Big[
    \frac{\lambda}{2}g(x(t))^2-g(x(t))F(x(t))-h(x(t))
  \Big]
  dt
  \Big\} \,.
\label{eq:weightforFlambdag}
\end{equation}
This shows that even if the biasing resulting from the parameter $\lambda$ can be partly reabsorbed into a non-equilibrium force by changing $F(x)$ into $F(x)-\lambda g(x)$, the non-equilibrium Langevin dynamics defined by Eq.~(\ref{eq:Langevin:modified}) is not equivalent to the biased dynamics defined by the action
Eq.~(\ref{eq:S-lambda1}), because of the remaining exponential reweighting given in Eq.~(\ref{eq:weightforFlambdag}). This remains true even in the simpler case when $A$ is the integrated current (or position of the particle), corresponding to $h(x)=0$ and $g(x)=1$.
We will explain in the next sections how an effective non-equilibrium dynamics, equivalent to the biased dynamics defined by the action Eq.~(\ref{eq:S-lambda1}),  can however be defined using a transformation of the operator $\WW _\lambda$.

\subsection{Large-deviation principle at large time}
\label{sec:large-devi-princ_large-time}

We now turn to the study of the large-time and weak-noise scaling behaviour of the distributions at hand.
One first remarks from~(\ref{eq:Wlambda}) that the biased operator $\WW_\lambda$ does not preserve probability (at odds with $\WW$, $\langle-|$ is not a left eigenvector of $\WW_\lambda$ of eigenvalue $0$).
In fact, the Perron--Frobenius theorem ensures that the maximal eigenvalue $\varphi_{\epsilon}(\lambda)$ of $\WW_\lambda$ is real and unique. We now assume that this operator has a gap (this is the case in general if the force is confining or if the space is compact); this ensures that at large time one has
\begin{equation}
  \label{eq:large-t-behaviour_operator}
  \ee^{t\WW_\lambda} 
  \ \stackrel[t\to\infty]{}{\sim} \
  \ee^{t\,\varphi_\epsilon(\lambda)}\:|R\rangle\langle L|
\quad\qquad\text{with}\quad
\varphi_\epsilon(\lambda) = \max \operatorname{Sp}\: \WW_\lambda
\end{equation}
where $|L\rangle$ and $|R\rangle$ are the corresponding left and right eigenvectors of $\WW_\lambda$, normalised as $\langle L| R\rangle=1$ and $\langle -| R\rangle=1$.
Then, the formal solution $|\hat P(t)\rangle = \ee^{t\WW_\lambda}|P_\ii\rangle$ of the evolution equation $\partial_t\hat P=\WW_\lambda\hat P$ implies that
\begin{equation}
  \label{eq:large-t-behaviour_Phatxt}
  \hat P(x,\lambda,t)
  \ \stackrel[t\to\infty]{}{\asymp} \
  \ee^{t\,\varphi_\epsilon(\lambda)}\:R(x) \,.
\end{equation}
Integrating~(\ref{eq:Phatlambdadef}) over $x$, this implies that, at large time, the moment generating function behaves as
\begin{equation}
  \label{eq:largetsMGF}
  \big\langle \ee^{-\frac\lambda\epsilon A(t)}\big\rangle 
  \ \stackrel[t\to\infty]{}{\asymp} \
  \ee^{t\,\varphi_\epsilon(\lambda)} \,.
\end{equation}
This result is an instance of a Donsker--Varadhan~\cite{donsker_asymptotic_1975} large deviation function (LDF) scaling: it indicates that the scaled cumulant generating function (sCGF) $\Phi_\epsilon(\lambda,t)$ defined as
\begin{equation}
 \label{eq:defCGFPhi}
\big\langle \ee^{-\frac{\lambda}{\epsilon}{A(t)}}\big\rangle=\ee^{t\,\Phi_\epsilon(\lambda,t)}
\end{equation}
goes to a constant at large $t$: $\lim_{t\to\infty}\Phi_\epsilon(\lambda,t)=\varphi_\epsilon(\lambda)$.
In other words, all cumulants of the observable $A(t)$ behave linearly in $t$ at large $t$.

Such LDF scaling can be translated into a large-time behaviour of the distribution of $A$: integrating~(\ref{eq:Phatlambdadef}) over $x$ one gets from~(\ref{eq:defCGFPhi}) that
\begin{equation}
  \label{eq:relationsMGF_PAt}
  \ee^{t\,\Phi_\epsilon(\lambda,t)}  =  \int  \ee^{-\frac{\lambda}{\epsilon}{A}}\:   {P}(A,t) \:dA \,.
\end{equation}
Since the l.h.s.~behaves exponentially in $t$ at large $t$, this indicates that the distribution $P(A\simeq a t,t)$ obeys the following scaling 
\begin{equation}
  \label{eq:LDFscaling_PAt}
  P(A\simeq a t,t) 
  \ \stackrel[t\to\infty]{}{\asymp} \
  \ee^{t\,\pi_\epsilon(a)}
\qquad\text{with}\quad
\varphi_\epsilon(\lambda) = \sup_{a} \big\{\pi_\epsilon(a)-\tfrac\lambda\epsilon a \big\} \,.
\end{equation}
This is an example of Gärtner--Ellis~\cite{gartner_large_1977,ellis_large_1984} LDF principle, obtained here through a saddle-point analysis of the integral in~(\ref{eq:relationsMGF_PAt}).
It indicates that, in the scaling $A\simeq a t$, the distribution of $A$ concentrates exponentially around the most probable value(s) of $a$, located at the maxima of the function $\pi_\epsilon(a)$.
If $\pi_\epsilon(a)$ is a concave function of $a$, then one can invert the Legendre--Fenchel transformation appearing in~(\ref{eq:LDFscaling_PAt}) and obtain
\begin{equation}
  \label{eq:legendrerelationpiphi}
  \pi_\epsilon(a) = \inf_{\lambda} \big\{ \varphi_\epsilon(\lambda)-\tfrac\lambda\epsilon a \big\} \,.
\end{equation}
These two Legendre--Fenchel transformations describe the change of ensemble between the microcanonical (fixed $a$) and canonical (fixed $\lambda$) descriptions, at fixed $\epsilon$.
The correspondence~(\ref{eq:legendrerelationpiphi}) can be extended at the level of trajectories: under the same convexity hypothesis, the conditioned distribution~\eqref{eq:defPAxtf} and the biased distribution~\eqref{eq:Phatlambdadef} are asymptotically equivalent~\cite{chetrite_nonequilibrium_2013,chetrite_nonequilibrium_2015} as $\tf\to\infty$, provided that the value of $\lambda$ is the one which realises the infimum in~(\ref{eq:legendrerelationpiphi}).

\subsection{Large-deviation principle in the weak-noise asymptotics $\epsilon \to 0$}
\label{sec:large-devi-princ_weak-noise}

We now consider the opposite order of limits, by keeping the duration $\tf$ finite and sending first the noise amplitude to $0$.
One can use the path-integral representation~(\ref{eq:Phatlambdapathintegral}) in order to study the weak-noise asymptotics of the distributions. By a saddle-point {evaluation} in the  $\epsilon\to 0$ limit, one sees from the definition~\eqref{eq:defCGFPhi} that, integrating~(\ref{eq:Phatlambdapathintegral}) over~$x$, the scaled CGF behaves as
\begin{equation}
  \Phi_\epsilon(\lambda,\tf) 
  \ \stackrel[\epsilon\to 0]{}{\sim} \
  \frac 1\epsilon \phi(\lambda,\tf)
\qquad\text{with}\quad
   \phi(\lambda,\tf) = -\frac{1}{\tf}\: \inf_{x(t){\vphantom{I_{I_{I_I}}}}} \Sla[x(t),\tf]
\label{eq:optimprincipphi-finitetf}
\end{equation}
where the action $\Sla[x(t),\tf]$ is defined in~(\ref{eq:S-lambda1}).
The optimisation is performed over trajectories $[x(t)]_{0\leq t\leq\tf}$ of duration $\tf$, whose initial position $x(0)$ is sampled according to the initial distribution $P_\ii$ (the simplest case is when $x(0)$ takes a fixed value), and whose final position $x(\tf)$ is optimised over in the $\inf$ of~(\ref{eq:optimprincipphi-finitetf}).
The function $\phi(\lambda,\tf)$ is a scaled CGF.
Since $\Phi_\epsilon(\lambda,\tf)$ converges to $\varphi_\epsilon(\lambda)$ as $\tf\to\infty$ for all $\epsilon$ (see~(\ref{eq:largetsMGF})), one expects that 
\begin{equation}
 \lim_{\tf\rightarrow\infty}\phi(\lambda,\tf) = \phi(\lambda)
\qquad\text{with}\quad
    \varphi_\epsilon(\lambda)   
    \ \stackrel[\epsilon\to 0]{}{\sim} \
    \frac 1 \epsilon \phi(\lambda) \,.
\label{eq:phiphiepsilon}
\end{equation}
In all, the saddle-point asymptotics provides the following optimisation principle for the sCGF $\phi(\lambda)$ as:
\begin{equation}
\label{eq:defvarphilambda}
 \phi(\lambda)=-\lim_{\tf\rightarrow\infty}\Big\lbrace\frac{1}{\tf}\inf_{x(t)} \Sla[x(t),\tf]\Big\rbrace \, .
\end{equation}
Hence, the determination of $\phi(\lambda)$ requires the knowledge of the optimal trajectories in the weak-noise limit. This is the topic of the next section.

\medskip

It is not obvious that the large-time and the weak-noise commute, \emph{i.e.}~that the sCGF $\phi(\lambda)$ given in~(\ref{eq:defvarphilambda}) by first taking $\epsilon\to0$ and then $\tf\to\infty$  is the same as the $\epsilon\to 0$ asymptotics $\phi(\lambda)=\lim_{\epsilon\to0} [\epsilon\,\varphi_\epsilon(\lambda)]$ (see~\eqref{eq:phiphiepsilon}) of the CGF $\varphi_\epsilon(\lambda)$ obtained from spectral considerations by first taking the $\tf\to\infty$ limit, as done in~(\ref{eq:largetsMGF}).
We will show in Sections~\ref{sec:case-peri-syst} and~\ref{sec:effective-dyn} that these two definitions coincide for periodic systems, \emph{i.e.}~that one can take the large-time and the weak-noise limits in whichever order one prefers.

\section{Determination of the sCGF $\phi(\lambda)$ for spatially periodic systems}
\label{sec:case-peri-syst}

\subsection{Optimal trajectories in the weak-noise limit}

We now aim at computing the scaled cumulant generating function $\phi(\lambda)$ by minimising the action $\Sla[x(t),\tf]$ according to Eq.~(\ref{eq:defvarphilambda}). 
The saddle-point equation for the optimal path sustaining a given fluctuation is obtained from the optimisation principle~(\ref{eq:optimprincipphi-finitetf}) and reads
\begin{equation}
 \label{eq:saddle-point-eq-x-F}
 \ddot{x}-F(x)F'(x)-\lambda h'(x)=0\:,
\end{equation}
{where the prime denotes a derivative with respect to $x$.}
We note that it does not depend on the function $g(x)$ since the term $\dot x(t)\,g(x(t))$ in the integrand of the action~(\ref{eq:S-lambda1}) is a total derivative.
It represents the conservative dynamics of a particle of unit mass in a potential
\begin{equation}
 \label{eq:devpotV}
 \mathcal V(x) = - \frac 12 F(x)^2 - \lambda h(x) \,.
\end{equation}
As a result, the energy
\begin{equation}
  \label{eq:defenergyxF}
  \mathcal E(\dot x,x) =
\frac{1}{2} \dot{x}^2 + \mathcal V(x)
\end{equation}
is conserved along an optimal trajectory.
Let us recall, however, that while optimal trajectories obey a deterministic conservative dynamics in the potential $\mathcal{V}(x)$ given by Eq.~(\ref{eq:devpotV}), the original dynamics of the problem obeys an overdamped Langevin dynamics, with a deterministic force $F(x)$ that may derive or not from a potential
---~see Eq.~(\ref{eq:Langevinepsilon}).  
Note that since $x(t)$ satisfies the second-order differential equation~\eqref{eq:saddle-point-eq-x-F}, it is uniquely specified in general only when one specifies a set of two parameters. It is convenient to choose for these two parameters the energy $\mathcal{E}$ and the initial position $x_{\rm i}$ (and the sign of the initial velocity, as we explain later).

Importantly, the initial position $x_{\rm i}$ for the optimal trajectory (solution of Eq.~(\ref{eq:saddle-point-eq-x-F})) differs from the `physical' initial condition $x(0)$ considered in the path integral~(\ref{eq:defPAxtf}) and sampled with $P_\ii$. Indeed, the underlying Langevin dynamics is dissipative, so that the biased distribution $\hat P(x,\lambda,\tf)$
converges to a steady state at large finite time $\tf$; then, the optimal trajectory that obeys the non-dissipative evolution~(\ref{eq:saddle-point-eq-x-F}) describes the most probable loci of $\hat P(x,\lambda,\tf)$\footnote{%
For instance, if $\hat P(x,\lambda,\tf)$ becomes peaked around a single point, the optimal trajectory solution of~(\ref{eq:saddle-point-eq-x-F}) is equal to that point, while if $\hat P(x,\lambda,\tf)$ becomes peaked around a contour, this optimal trajectory follows that contour  ---~as usual in the WKB description.
}%
.
In other words, there is a transient regime for the  `physical' initial distribution $P_\ii(x)$ to reach a distribution $\hat P(x,\lambda,\tf)$ that falls into a consistent weak-noise description.
As a result, the initial distribution $P_\ii(x)$ becomes irrelevant after this transient regime and can thus be forgotten in the weak-noise evaluation of the path-integral~(\ref{eq:defPAxtf}).

On the other hand, Eq.~(\ref{eq:saddle-point-eq-x-F}) has an infinite number of solutions, only one of them being the actual optimal trajectory that minimises the action $\Sla[x(t),\tf]$.
To determine this optimal trajectory, one has to parametrise each solution of Eq.~(\ref{eq:saddle-point-eq-x-F}) by the initial position $x_{\rm i}$ and the energy $\mathcal{E}$, but again, $x_{\rm i}$ differs from the `physical' initial condition $x(0)$.

In the following, we consider finite-size spatially periodic systems, for which $F(x+1)=F(x)$, $h(x+1)=h(x)$ and $g(x+1)=g(x)$ (we took the spatial period as the unit length, without loss of generality).
For such systems, the optimal trajectory minimising the action $\Sla[x(t),\tf]$ becomes independent of $x(0)$ for large enough time~$\tf$, so that the only relevant parameter to characterise the trajectories in this limit is their energy. 
To determine the sCGF $\phi(\lambda)$, one has to evaluate the action $\Sla[x(t),\tf]$ for any of the optimal trajectories given by Eq.~(\ref{eq:saddle-point-eq-x-F}), and to find the optimal trajectory that minimises the action. In practice, this last step consists in minimising the action over the energy of the trajectories.
Using Eqs.~(\ref{eq:S-lambda1}) and~(\ref{eq:devpotV}), the action 
$\Sla[x(t),\tf]$ can be written as
\begin{equation}
\Sla[x(t),\tf] = \int_0^{\tf} dt \, \mathcal{L}(\dot{x}(t), x(t))
\label{eq:SlaL}
\end{equation}
with a Lagrangian
\begin{equation}
\label{eq:def:Lagrangian}
\mathcal{L}(\dot x, x) = \frac{1}{2} \dot{x}^2 - \mathcal{V}(x)
+ \dot{x} \big( \lambda g(x) - F(x) \big) \,.
\end{equation}
Note that the last term, proportional to $\dot{x}$, in Eq.~(\ref{eq:def:Lagrangian}) plays no role in Eq.~(\ref{eq:saddle-point-eq-x-F}) since it is a total derivative, but it has to be included in the Lagrangian to correctly evaluate (and minimise) the action.

Assuming that the force field $F(x)$ and the function $h(x)$ are bounded, the potential $\mathcal{V}(x)$ is also bounded. We denote as $\mathcal{V}_{\rm max}$ the maximum value of the potential:
\begin{equation} \label{eq:def:Vmax}
\mathcal{V}_{\rm max} = \max_x \mathcal{V}(x) \,.
\end{equation}
The value $\mathcal{V}_{\rm max}$ allows one to classify the optimal trajectories $\xstar(t)$ into periodic and propagative solutions, according to their energy $\mathcal{E}$ (for convenience, we include constant trajectories as a special case of the periodic ones). For $\mathcal{E}<\mathcal{V}_{\rm max}$, optimal trajectories are confined by the potential $\mathcal V(x)$, and are periodic in time.
For $\mathcal{E}>\mathcal{V}_{\rm max}$, the potential no longer confines the optimal trajectories, which are then propagative, with a constant sign of the velocity $\dot{x}$.
As a result, the sCGF $\phi(\lambda)$ is obtained by minimising the action over both sets of periodic and propagative optimal trajectories.
One can thus write, taking into account the minus sign in Eq.~(\ref{eq:defvarphilambda}),
\begin{equation} \label{eq:phi:lambda:max:per:prop}
\phi(\lambda) = \max \left\{ \phi_{\rm per}(\lambda),  \phi_{\rm prop}(\lambda) \right\}
\end{equation}
where $\phi_{\rm per}(\lambda)$, $\phi_{\rm prop}(\lambda)$ are defined by minimising the action over the sets of periodic and propagative trajectories respectively:
\begin{align}
\label{eq:def:phi:per}
 \phi_{\rm per}(\lambda) &= -\lim_{\tf\rightarrow\infty}\Big\lbrace\frac{1}{\tf}
\inf_{\mathcal{E}<\mathcal{V}_{\rm max}} \Sla[\xstar(t),\tf]\Big\rbrace\,\,, \\
\label{eq:def:phi:prop}
 \phi_{\rm prop}(\lambda) &= -\lim_{\tf\rightarrow\infty}\Big\lbrace\frac{1}{\tf}
\inf_{\mathcal{E}>\mathcal{V}_{\rm max}} \Sla[\xstar(t),\tf]\Big\rbrace\,\,.
\end{align}
In the following, we successively evaluate $\phi_{\rm per}(\lambda)$ and $\phi_{\rm prop}(\lambda)$.

\subsection{Time-periodic trajectories}
\label{sec:time-per-traj}

We start by evaluating $\phi_{\rm per}(\lambda)$.
A particular type of periodic trajectories are the time-independent ones, for which $\dot\xstar=0$ and $\xstar=x_0$, implying $\mathcal{V}'(x_0)=0$ from Eq.~(\ref{eq:saddle-point-eq-x-F}).
For such trajectories, one has $\mathcal{L}(\dot \xstar, \xstar)=-\mathcal{V}(x_0)$, so that
minimising the action over time-independent trajectories selects points $x_0$ that are at the location(s) of the maximum of the potential $\mathcal{V}(x)$; hence:
\begin{equation}
 \lim_{\tf\rightarrow\infty}\Big\lbrace\frac{1}{\tf}\inf_{x(t)=x_0} \Sla[\xstar(t),\tf]\Big\rbrace = -\mathcal{V}_{\rm max}.
\end{equation}

Considering now a generic time-periodic optimal trajectory, the action reads,
with $\xstar \equiv \xstar(t)$
\begin{equation}\label{eq:action:per:traj}
\fl\qquad\quad
\frac{1}{\tf} \Sla[\xstar(t),\tf] = \underbrace{\frac{1}{2\tf} \int_0^{\tf} dt \, (\dot \xstar)^{2}}_{\ge 0} \; 
\underbrace{-\frac{1}{\tf} \int_0^{\tf} dt \, \mathcal{V}\big(\xstar\big)}_{\ge -\mathcal{V}_{\rm max}}
+ \underbrace{\frac{1}{\tf} \int_0^{\tf} dt \, \dot \xstar \Big( \lambda g\big(\xstar\big) - F\big(\xstar\big) \Big)}_{
 \rightarrow \, 0 \ \text{when} \ \tf \to \infty
}
\:.
\end{equation}
The proof that the last integral in~(\ref{eq:action:per:traj}) goes to $0$ when $\tf \to \infty$ comes from a change of variable from $t$ to $x$:
\begin{equation}
 \frac{1}{\tf} \int_0^{\tf} dt \, \dot \xstar \Big( \lambda g(\xstar) - F(\xstar) \Big)
= \frac{1}{\tf} \int_{\xstar(0)}^{\xstar(\tf)} dx \Big( \lambda g(x) - F(x) \Big)
\end{equation}
which goes to $0$ when $\tf \to \infty$ because $[\lambda g(x) - F(x)]$ is bounded on the finite interval $[{\xstar(0)},{\xstar(\tf)}]$.
It is tempting to take the $\tf\to\infty$ limit and to conclude from~(\ref{eq:def:phi:per}) that $\phi_{\rm per}(\lambda)\leq \mathcal{V}_{\rm max}$
 but this would require to exchange the $\inf$ and the $\tf\to\infty$ limit in~(\ref{eq:def:phi:per}), which enters in conflict with our goal since we are interested in how the small-noise and large-time limits commute.
To avoid this exchange, one writes from~(\ref{eq:action:per:traj}) that for any time-periodic optimal trajectory,
\begin{equation}\label{eq:timeper:action0}
\frac{1}{\tf} 
\inf_{\mathcal{E}<\mathcal{V}_{\rm max}}
\Sla[\xstar(t),\tf]
\ge -\mathcal{V}_{\rm max} 
-
\frac{1}{\tf} \inf_{\mathcal{E}<\mathcal{V}_{\rm max}} \int_{\xstar(0)}^{\xstar(\tf)} dx \Big( \lambda g(x) - F(x) \Big)
\end{equation}
so that taking the $\tf \to \infty$ limit one finds 
\begin{equation}\label{eq:timeper:action}
\lim_{\tf\rightarrow\infty} \Big\lbrace\frac{1}{\tf} \inf_{\mathcal{E}<\mathcal{V}_{\rm max}} \Sla[\xstar(t),\tf]\Big\rbrace
\ge -\mathcal{V}_{\rm max} \,,
\end{equation}
because the integrand on the r.h.s.~of~(\ref{eq:timeper:action0}) is a bounded function on an interval of fixed finite length.
From the definition~(\ref{eq:def:phi:per}), we obtain $\phi_{\rm per}(\lambda)\leq \mathcal{V}_{\rm max}$.
Remarking now from~(\ref{eq:SlaL}-\ref{eq:def:Lagrangian}) that this bound is realised for time-independent trajectories $\xstar=x_0$ we conclude that
\begin{equation}
\label{eq:phi:lambda:per:Vmax}
 \phi_{\rm per}(\lambda) = \mathcal{V}_{\rm max}(\lambda)
\end{equation}
where the $\lambda$-dependence of $\mathcal{V}_{\rm max}$ has been made explicit. Therefore, for $\mathcal{E}<\mathcal{V}_{\rm max}$, the optimal trajectories sustaining a given fluctuations are time-independent of the form $\xstar=x_0$, where $x_0$ are the points maximising the potential $\mathcal{V}(x_0)=\mathcal{V}_{\rm max}$.

\subsection{Propagative trajectories}
\label{sec:propagative-trajs}

To evaluate $\phi_{\rm prop}(\lambda)$,
one now has to compute the minimum of the action over all propagative optimal trajectories, \emph{i.e.}, trajectories for which $\mathcal{E} > \mathcal{V}_{\rm max}$.
Then, from energy conservation, one has
\begin{equation}
  \label{eq:xdotofx}
  \dot \xstar = \sigma \sqrt{2\big(\mathcal E - \mathcal{V}(\xstar)\big)}
\end{equation}
where $\sigma = \pm 1$ is the sign of $\dot \xstar$
(we recall that the sign of $\dot \xstar$ is constant all along propagative trajectories).
Propagative trajectories are pseudo-periodic, in the sense that
$\xstar(t+T)=\xstar(t)+\sigma$, which may be identified with $\xstar(t)$ due to the spatial periodicity of the system; $T=T(\mathcal{E})$ is the pseudo-period $T(\mathcal{E})$, determined as
\begin{equation}
  \label{eq:expressionperiodT}
  T = \int_0^T dt = \int_0^1\frac{dx}{\sqrt{2\big( \mathcal{E} - \mathcal{V}(x)\big)}}
  \:,
\end{equation}
where we have used Eq.~(\ref{eq:xdotofx}) to change the integration variable from $t$ to $x$.
Using the relation
\begin{equation}
\mathcal{L}(\dot \xstar,\xstar) = \mathcal{E}-2\mathcal{V}(\xstar) + \dot \xstar \big( \lambda g(\xstar) - F(\xstar) \big) \,,
\end{equation}
the action of a propagative optimal trajectory on the time interval $[0,T(\mathcal{E})]$ is, expanding the Lagrangian~(\ref{eq:def:Lagrangian}),
\begin{equation}
  \label{eq:evalactionopt}
  S_\lambda[\xstar(t),T(\mathcal{E})] 
= \sigma \int_0^1  \big( \lambda g(x) - F(x) \big) dx + T(\mathcal{E}) \mathcal{E}
- \int_0^1  \frac{2\mathcal{V}(x)}{\sqrt{2\big( \mathcal{E} - \mathcal{V}(x)\big)}}\:dx \,.
\end{equation}
To lighten notations, we define
\begin{equation}
B = \int_0^1  \big( \lambda g(x) - F(x) \big) dx \,, \qquad
R(\mathcal{E}) = \int_0^1  \frac{2\mathcal{V}(x)}{\sqrt{2\big( \mathcal{E} - \mathcal{V}(x)\big)}}\:dx \,.
 \label{eq:def:B:K}
\end{equation}
Note that the term $F(x)$ in the integral defining $B$ gives no contribution when the force $F(x)$ derives from a potential.

In the large-time limit, the value of the action over every interval $[n T(\mathcal{E}), (n+1)T(\mathcal{E})]$ is the same (by periodicity of the optimal trajectory). Furthermore, the optimal trajectory dependence on the initial value $x_0$ is now replaced by a pseudo-periodic boundary condition of the form {$x(1)=x(0)+\sigma$}. Hence the $\phi_{\rm prop}(\lambda)$ defined in Eq.~\eqref{eq:def:phi:prop} is equal to:
\begin{align}
\label{eq:varphiperF}
\phi_{\rm prop}(\lambda)
&= -\lim_{n \rightarrow\infty}\; \inf_{\mathcal{E} > \mathcal{V}_{\rm max},\, \sigma=\pm 1}  \left\lbrace \frac{1}{n T(\mathcal{E})}  \int_0^{nT(\mathcal{E})}  \mathcal L(\dot x^\star,x^\star)\:dt \right\rbrace
\label{eq:phiasminoverE0}
\\
  & = - \inf_{\mathcal{E} > \mathcal{V}_{\rm max},\, \sigma=\pm 1} \Big\{ \frac{1}{T(\mathcal{E})} \int_0^{T(\mathcal{E})} \mathcal L(\dot x^\star,x^\star)\:dt \Big\}
\label{eq:phiasminoverE}
\\
  & = - \inf_{\mathcal{E} > \mathcal{V}_{\rm max},\, \sigma=\pm 1} \Big\{ \mathcal{E} + \frac{\sigma B - R(\mathcal{E})}{T(\mathcal{E})} \Big\}
\label{eq:varphiperF2}
\end{align}
where in~(\ref{eq:phiasminoverE0})-(\ref{eq:phiasminoverE}) the optimal trajectory $\xstar(t)$ is the propagative solution of the saddle-point equation, with an energy $\mathcal{E}$ and a pseudo-period $T(\mathcal E)$ that depends on $\mathcal E$, as inferred from~(\ref{eq:expressionperiodT}).
Determining $\phi_{\rm prop}(\lambda)$ thus amounts to finding, for both $\sigma=\pm 1$, the infimum of the function
\begin{equation}
\Psi_{\sigma}(\mathcal{E}) = \mathcal{E} + \frac{\sigma B - R(\mathcal{E})}{T(\mathcal{E})} \,.
\label{eq:defCapitalPsi}
\end{equation}
The function $\Psi_{\sigma}(\mathcal{E})$ is defined over the interval
$(\mathcal{V}_{\rm max},+\infty)$.
When $\mathcal{E}\rightarrow \mathcal{V}_{\rm max}$, both $R(\mathcal{E})$ and
$T(\mathcal{E})$ diverge to infinity (assuming $\mathcal{V}(x)$ is regular close to $\mathcal{V}_{\rm max}$), but their ratio
$R(\mathcal{E})/T(\mathcal{E})\rightarrow 2\mathcal{V}_{\rm max}$, so that
$\Psi_{\sigma}(\mathcal{E}) \rightarrow -\mathcal{V}_{\rm max}$.
In the opposite limit $\mathcal{E} \rightarrow \infty$,
$T(\mathcal{E}) \sim R(\mathcal{E}) \sim 1/\sqrt{\mathcal{E}}$,
yielding $\Psi_{\sigma}(\mathcal{E}) \rightarrow +\infty$.
Consequently, if $\Psi_{\sigma}(\mathcal{E})$ has no minimum for $\mathcal{E}\in(\mathcal{V}_{\rm max},+\infty)$, one has:
\begin{equation}
\inf_{\mathcal{E} > \mathcal{V}_{\rm max}} \Psi_{\sigma}(\mathcal{E}) = -\mathcal{V}_{\rm max} \,.
\end{equation}
We now proceed to determine if $\Psi_{\sigma}(\mathcal{E})$ has a minimum $\mathcal{E}_{\sigma}^{\ast}$, satisfying $\Psi_{\sigma}'(\mathcal{E}_{\sigma}^{\ast})=0$.
The derivative $\Psi_{\sigma}'(\mathcal{E})$ reads
\begin{equation}
\Psi_{\sigma}'(\mathcal{E}) = \frac{1}{T(\mathcal{E})^2} \Big[ T(\mathcal{E})^2 - R'(\mathcal{E}) T(\mathcal{E}) + R(\mathcal{E}) T'(\mathcal{E}) -\sigma B T'(\mathcal{E}) \Big]\,.
\end{equation}
From the definition~(\ref{eq:def:B:K}) of $R(\mathcal{E})$, one finds that
\begin{equation} \label{eq:RTET}
R'(\mathcal{E}) = T(\mathcal{E}) + 2\mathcal{E} T'(\mathcal{E})
\end{equation}
so that $\Psi_{\sigma}'(\mathcal{E})$ can be rewritten as
\begin{equation}
\Psi_{\sigma}'(\mathcal{E}) = \frac{T'(\mathcal{E})}{T(\mathcal{E})^2}
\Big[ R(\mathcal{E}) - 2\mathcal{E} T(\mathcal{E}) -\sigma B \Big]\,.
\end{equation}
Since $T'(\mathcal{E}) \ne 0$ for all $\mathcal{E}$, the condition $\Psi_{\sigma}'(\mathcal{E}_{\sigma}^{\ast})=0$ is equivalent to
\begin{equation} \label{eq:determin:Estar}
R(\mathcal{E}_{\sigma}^{\ast}) - 2\mathcal{E}_{\sigma}^{\ast} T(\mathcal{E}_{\sigma}^{\ast}) -\sigma B = 0
\end{equation}
which determines $\mathcal{E}_{\sigma}^{\ast}$. 
If a solution $\mathcal{E}_{\sigma}^{\ast}$ exists, one has from Eqs.~(\ref{eq:defCapitalPsi}) and~(\ref{eq:determin:Estar})
\begin{equation}
\Psi_{\sigma}(\mathcal{E}_{\sigma}^{\ast})
= \mathcal{E}_{\sigma}^{\ast} + \frac{\sigma B - R(\mathcal{E}_{\sigma}^{\ast})}{T(\mathcal{E}_{\sigma}^{\ast})} = - \mathcal{E}_{\sigma}^{\ast} \,.
\label{eq:expressionPsisigma}
\end{equation}
Using Eqs.~(\ref{eq:RTET}) and~(\ref{eq:determin:Estar}), one can show that
the second derivative $\Psi_{\sigma}''(\mathcal{E}_{\sigma}^{\ast})$ takes the simple form
\begin{equation}
\Psi_{\sigma}''(\mathcal{E}_{\sigma}^{\ast}) = -\frac{T'(\mathcal{E}_{\sigma}^{\ast})}{T(\mathcal{E}_{\sigma}^{\ast})} = \frac{\int_0^1 \big( \mathcal{E}_{\sigma}^{\ast}-\mathcal{V}(x)\big)^{-3/2}\, dx}{2\int_0^1 \big( \mathcal{E}_{\sigma}^{\ast}-\mathcal{V}(x)\big)^{-1/2}\, dx} >0 \,,
\end{equation}
so that $\mathcal{E}_{\sigma}^{\ast}$ is a local minimum. The fact that it is a global minimum comes from a unicity argument, which goes as follows.
Eq.~(\ref{eq:determin:Estar}) can be rewritten using
Eqs.~(\ref{eq:expressionperiodT}) and~(\ref{eq:def:B:K}) as
\begin{equation} \label{eq:determin:Estar2}
\int_0^1 \sqrt{2\big( \mathcal{E}_{\sigma}^{\ast}-\mathcal{V}(x)\big)}\, dx = -\sigma B \,.
\end{equation}
The integral on the l.h.s.~of Eq.~(\ref{eq:determin:Estar2}) spans the interval
$(\int_0^1 \sqrt{2(\mathcal{V}_{\rm max}-\mathcal{V}(x))}\,dx, +\infty)$ as a function of $\mathcal{E}_{\sigma}^{\ast}$. Hence a solution $\mathcal{E}_{\sigma}^{\ast}$ exists if
\begin{equation} \label{eq:condit:B}
- \sigma B > \int_0^1 \sqrt{2\big(\mathcal{V}_{\rm max}-\mathcal{V}(x)\big)}\,dx
\end{equation}
where we recall that $B$ is defined in Eq.~(\ref{eq:def:B:K}).
Since $\int_0^1 \sqrt{2(\mathcal{E}-\mathcal{V}(x))}\, dx$ is an increasing function of $\mathcal{E}$, the solution $\mathcal{E}_{\sigma}^{\ast}$ is unique if it exists.
Hence the function $\Psi_{\sigma}(\mathcal{E})$ has at most one stationary point, so that its local minimum $\mathcal{E}_{\sigma}^{\ast}$ is, if it exists, a global minimum.
In addition, if $\mathcal{E}_{\sigma}^{\ast}$ exists, then 
$\mathcal{E}_{-\sigma}^{\ast}$ does not exist, since Eq.~(\ref{eq:condit:B}) cannot be simultaneously satisfied for $\sigma$ and $-\sigma$.
This implies that $\mathcal{E}_{\sigma}^{\ast}$ can exist only for
\begin{equation} \label{eq:sigma:signB}
\sigma=-{\rm sign}(B)\,,
\end{equation}
and Eq.~(\ref{eq:condit:B}) can be rewritten as
\begin{equation} \label{eq:condit:B2}
|B| > \int_0^1 \sqrt{2\big(\mathcal{V}_{\rm max}-\mathcal{V}(x)\big)}\,dx \,.
\end{equation}
If Eq.~(\ref{eq:condit:B2}) is satisfied, one has for $\sigma=-{\rm sign}(B)$ that
\begin{equation}
\inf_{\mathcal{E} > \mathcal{V}_{\rm max}} \Psi_{\sigma}(\mathcal{E}) = -\mathcal{E}_{\sigma}^{\ast}, \quad \text{and} \quad
\inf_{\mathcal{E} > \mathcal{V}_{\rm max}} \Psi_{-\sigma}(\mathcal{E}) = -\mathcal{V}_{\rm max} \,.
\end{equation}
As a result, Eq.~(\ref{eq:varphiperF2}) implies
\begin{equation} \label{eq:phi:lambda:prop:E}
\phi_{\rm prop}(\lambda) = -\min \{-\mathcal{E}_{\sigma}^{\ast},-\mathcal{V}_{\rm max}\}
= \max\{\mathcal{E}_{\sigma}^{\ast},\mathcal{V}_{\rm max}\}
= \mathcal{E}_{\sigma}^{\ast}
\end{equation}
(with $\sigma=-{\rm sign}(B)$) if Eq.~(\ref{eq:condit:B2}) holds.

\medskip
In the opposite case, if Eq.~(\ref{eq:condit:B2}) is not satisfied,
\begin{equation}
\inf_{\mathcal{E} > \mathcal{V}_{\rm max}} \Psi_{\sigma}(\mathcal{E}) =
\inf_{\mathcal{E} > \mathcal{V}_{\rm max}} \Psi_{-\sigma}(\mathcal{E}) =
-\mathcal{V}_{\rm max} \,,
\end{equation}
so that
\begin{equation} \label{eq:phi:lambda:prop:Vmax}
\phi_{\rm prop}(\lambda) = \mathcal{V}_{\rm max} \,.
\end{equation}
Since from Eq.~(\ref{eq:phi:lambda:per:Vmax})
$\phi_{\rm per}(\lambda)= \mathcal{V}_{\rm max} \le \phi_{\rm prop}(\lambda)$, 
Eq.~(\ref{eq:phi:lambda:max:per:prop}) implies that for all $\lambda$
\begin{equation} \label{eq:phi:phiprop}
\phi(\lambda) = \phi_{\rm prop}(\lambda) \,.
\end{equation}

\bigskip

To summarise, we have shown that when a propagative optimal solution exists, it is unique and the value of the corresponding sCGF is given by the \emph{energy} $\mathcal{E}_{\sigma}^{\ast}$ of such trajectory; otherwise, the value of the sCGF is given by the maximum value $\mathcal{V}_{\rm max}$ of the effective potential $\mathcal{V}(x)$.
Note that the criterion given by Eq.~(\ref{eq:condit:B2}) for the existence of an optimal propagative solution can be interpreted as a condition on $\lambda$; using explicit notations, one has:
\begin{equation} \label{eq:cdtn:lambda:exist:Estar}
  \fl \qquad\quad
\left| \lambda \int_0^1 g(x)\,dx - \int_0^1 F(x)\,dx \right|
> \int_0^1 \sqrt{2\mathcal{V}_{\rm max}(\lambda)+ F(x)^2 +2\lambda h(x) }\; dx 
\quad
\Rightarrow
\quad
\begin{array} {l} \textnormal{\small{optimal traj.}}\\ \textnormal{\small{is propagative}}\end{array}
\:.
\end{equation}

\begin{figure}[t]
\begin{center}
  \includegraphics[width=.666666\columnwidth]{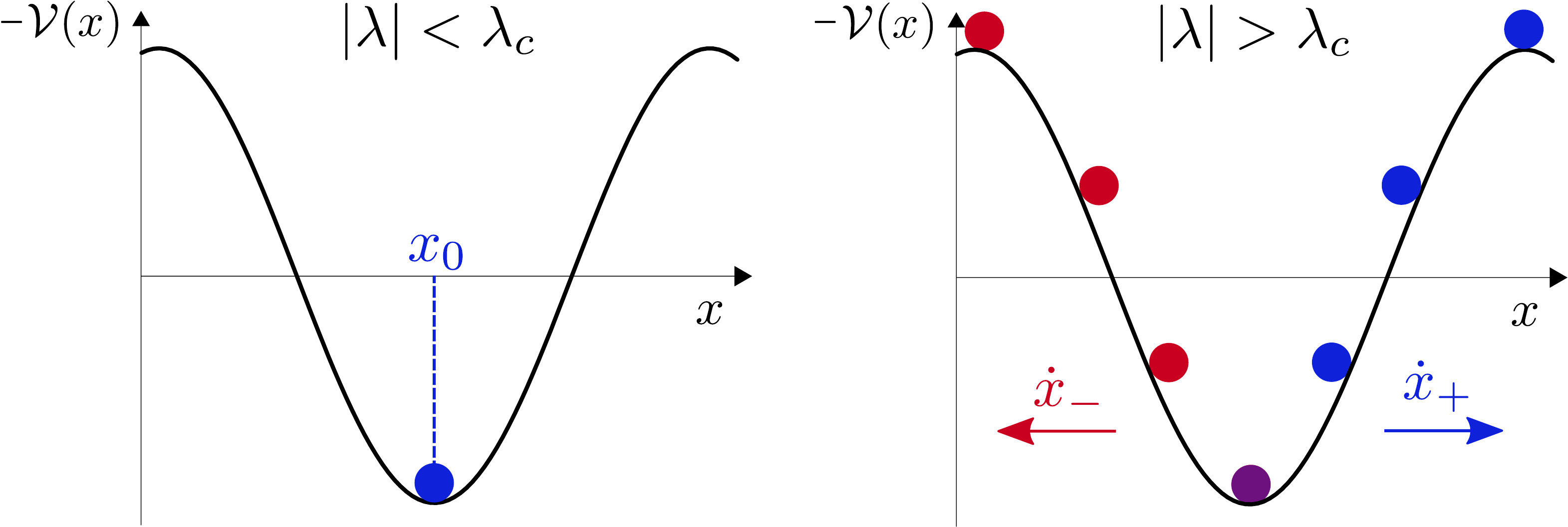}
\end{center}
  \caption{The different classes of optimal trajectories of interest: On the left, stationary ones, with $x_0$ the location of the maximum of $\mathcal V(x)$, defined in Eq.~(\ref{eq:devpotV}); On the right, propagative ones, either increasing or decreasing in time.
Periodic trajectories which oscillate around $x_0$ without being propagative have a larger action than the stationary one in $x_0$, as shown in Subsection~\ref{sec:time-per-traj}.
The criterion for the existence of propagative trajectories is given by Eq.~(\ref{eq:cdtn:lambda:exist:Estar}).
\label{fig:sketch-3phases-2}
}
\end{figure}

\medskip
In conclusion, the evaluation of $\phi(\lambda)$ generically goes as follows. For each value of $\lambda$, one checks whether Eq.~(\ref{eq:cdtn:lambda:exist:Estar}) is satisfied. If it holds, one determines $\mathcal{E}_{\sigma}^{\ast}$ by solving Eq.~(\ref{eq:determin:Estar}) with $\sigma=-{\rm sign}(B)$, \emph{i.e.},
\begin{equation} \label{eq:determin:Estar2bis}
R(\mathcal{E}_{\sigma}^{\ast}) - 2\mathcal{E}_{\sigma}^{\ast} T(\mathcal{E}_{\sigma}^{\ast}) + |B| = 0 \,,
\end{equation}
leading to $\phi(\lambda)=\mathcal{E}_{\sigma}^{\ast}(\lambda)$.
If Eq.~(\ref{eq:cdtn:lambda:exist:Estar}) is not satisfied, then $\phi(\lambda)=\mathcal{V}_{\rm max}(\lambda)$.
This result allows one to determine the existence of possible dynamical phase transitions in the fluctuations of the additive observable $A(\tf)$.
When varying $\lambda$, one can indeed jump from a situation where the optimal trajectory is time-independent (when Eq.~(\ref{eq:cdtn:lambda:exist:Estar}) is not satisfied) to 
a situation where the optimal trajectory is time-dependent.
Such a transition between two classes of optimal trajectories  is illustrated in Fig.~\ref{fig:sketch-3phases-2}
and corresponds to a breaking of the `additivity principle'~\cite{bodineau_current_2004}).

For an activity-type observable $A(\tf)$ ($g(x)=0$) in the presence of a conservative force $F(x)=-U'(x)$, the l.h.s.~of Eq.~(\ref{eq:cdtn:lambda:exist:Estar}) is equal to $0$, so that Eq.~(\ref{eq:cdtn:lambda:exist:Estar}) is never satisfied. 
It follows that $\phi(\lambda)=\mathcal{V}_{\rm max}(\lambda)$ for all~$\lambda$, meaning that the optimal trajectory is always time-independent in this case (in other words, there is no breaking of the additivity principle).
This however does not forbid dynamical phase transitions since, as seen from the expression~(\ref{eq:devpotV}) of $\mathcal V(x)$ the `tilting' contribution $-\lambda h(x)$ can make the location of the maximum of $\mathcal V(x)$ switch from one position to another, if for instance $F(x)$ presents more than one equilibrium point.
Such a situation occurs for instance in the large deviation of additive observables in driven diffusive systems~\cite{PhysRevLett.118.030604}.

We discuss below the determination of $\phi(\lambda)$ in the case of current-type additive observable, which generically leads to a phase transition between stationary and non-stationary trajectories.

\subsection{Determination of $\phi(\lambda)$ for current-type additive observable ($h(x)=0$) and conservative force $F(x)$}
\label{sec:case-purely-current}

Considering a current-type additive observable (corresponding to $h(x)=0$)
as well as a conservative force $F(x)=-U'(x)$,
Eq.~(\ref{eq:cdtn:lambda:exist:Estar}) simplifies to
\begin{equation} \label{eq:cdtn:lambda:exist:Estar:current}
|\lambda| >  \lambda_\cc \equiv \frac{\int_0^1 |F(x)|\, dx}{\int_0^1 g(x)\,dx}
\end{equation}
where we have assumed that $\int_0^1 g(x)\,dx >0$ (the case $\int_0^1 g(x)\,dx <0$ is treated in the same way), and used the fact that
$\mathcal{V}_{\rm max}=0$ when $h(x)=0$ (as inferred from~(\ref{eq:devpotV}) and~(\ref{eq:def:Vmax})).
For $|\lambda| > \lambda_\cc$, $\phi(\lambda)$ is solution  of the equation
\begin{equation}
\int_0^1 \sqrt{2\phi(\lambda)+F(x)^2}\, dx = |\lambda| \int_0^1 g(x)\,dx
\label{eq:eqforphi-h0}
\end{equation}
while for $| \lambda| < \lambda_\cc$, $\phi(\lambda)=\mathcal{V}_{\rm max}=0$.
Consequently, $\lambda_\cc$ is the critical value at which the dynamical phase transition takes place.
Note that $\phi(\lambda)$ is an even function of $\lambda$ for current-type additive observable in a system with a conservative force, due to time-reversal invariance.

The singular behaviour of the sCGF close to the transition depends both on local and global properties of the force $F(x)$.
If one naively expands for small $\phi(\lambda)$
\begin{equation}
  \label{eq:devsqrt}
  \sqrt{F(x)^2+2\phi(\lambda)} = |F(x)| + \frac{\phi(\lambda)}{|F(x)|} + O\big(\phi(\lambda)^2\big)
\end{equation}
and integrates over $0<x<1$ in order to obtain an expansion of Eq.~\eqref{eq:eqforphi-h0} for $\phi(\lambda)$, one finds a divergent integral
$\int_0^1 dx/|F(x)|$ (if there is a fixed point $F(x_0)=0$ with $F'(x_0) \ne 0$, which is the case in general).
The expansion at small $\phi(\lambda)$ is thus ill-defined. The logarithmic divergence of $\int_0^1 dx/|F(x)|$  suggests a behaviour  $\phi(\lambda) \sim (\lambda-\lambda_\cc) / |\log(\lambda-\lambda_\cc)|$  but the situation is  better understood on a specific example first.

\medskip

\begin{figure}[t]
\begin{center}
  \includegraphics[width=.5\columnwidth]{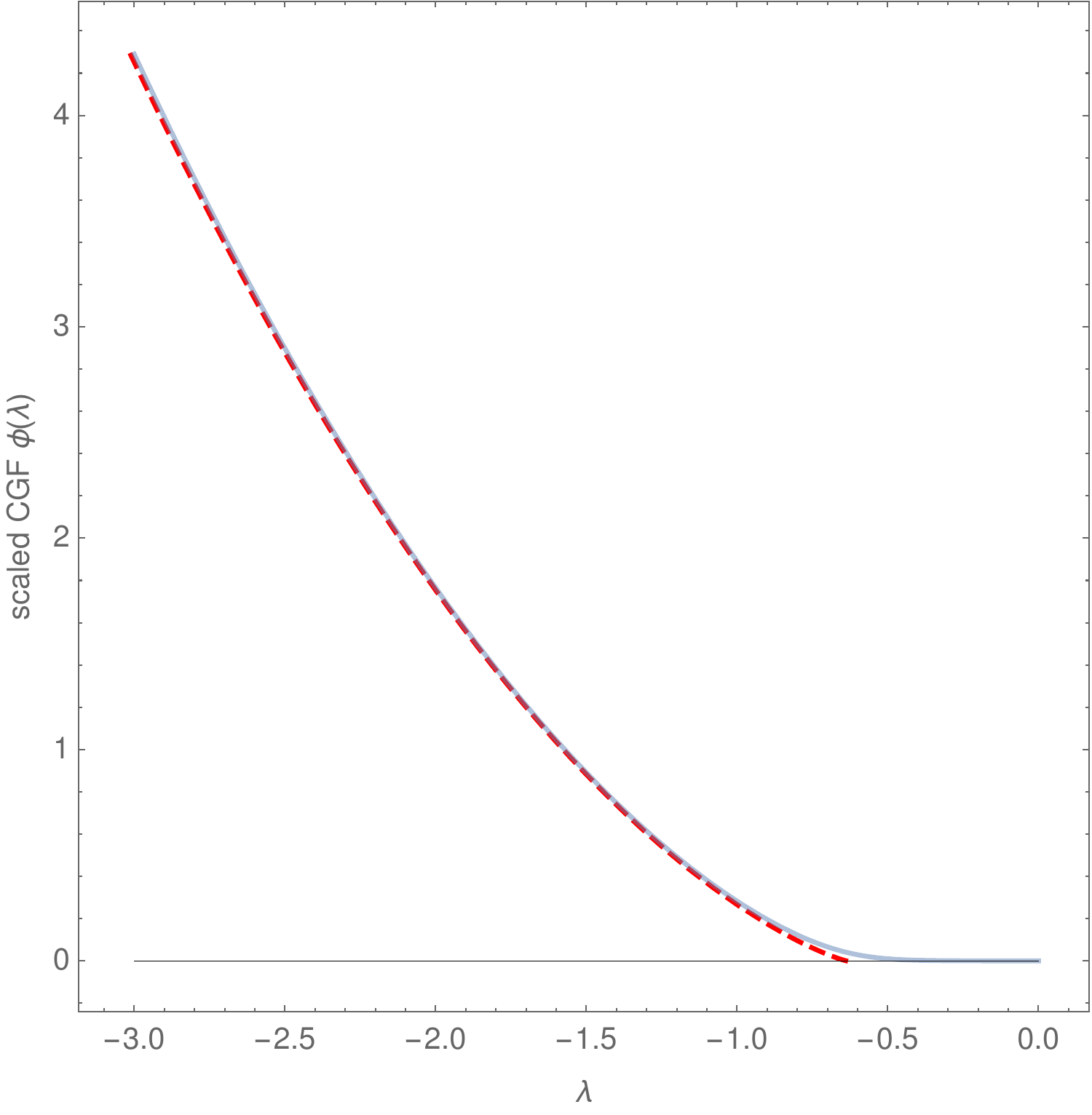}
\end{center}
  \caption{%
An example sCGF $\phi(\lambda)$, for $F(x)=\sin(2\pi x)$ and  $h(x)=0$ and $g(x)=1$.
Comparison of the evaluation of the scaled CGF $\phi(\lambda)$ between the weak-noise approach (deduced from~(\ref{eq:equationforE})-(\ref{eq:finalresultphioflambda}), dashed red line) and the maximal eigenvalue of the deformed operator~(\ref{eq:Wlambda}) of a lattice version of the dynamics (translucent blue line; $128$ sites, $\epsilon=0.075$).
In the negative regime of $\lambda$, the transition occurs at $\lambda=-\lambda_\cc=-\frac 2\pi\simeq-0.637$.
\label{fig:phioflambda_comparision-WKB-diag}
}
\end{figure}

As an explicit example, we consider the case $F(x)=\sin (2\pi x)$\,, $h(x)=0$ and $g(x)=1$, where the force $F(x)$ is conservative, and the observable $A(t)$ is the integrated current (or the position at time $t$, counted with the number of turns).
Eq.~(\ref{eq:cdtn:lambda:exist:Estar:current}) straightforwardly leads to
\begin{equation}
  \label{eq:lambdacFsin}
  \lambda_\cc = \frac 2\pi \,.
\end{equation}
For $|\lambda|>\lambda_\cc$, the scaled CGF $\phi(\lambda)$ is solution in $\mathcal E_\lambda$ of the equation
\begin{equation}
  \label{eq:equationforE}
   \lambda = \cases{ - \Lambda(\mathcal E_\lambda) \quad \text{for}\,\lambda<-\lambda_\cc 
   \\
    \Lambda(\mathcal E_\lambda) \quad \text{for}\,\lambda>\lambda_\cc \,,
   }
   \end{equation}
with
\begin{equation}
   \Lambda(\mathcal E) 
   \equiv 
   \frac 2 \pi \sqrt{2\mathcal E}\,
   \operatorname E\!\left(-\frac{1}{{2\mathcal E}}\right) ,
\end{equation}
where $ \operatorname E(\cdot)$ is the complete elliptic integral of the second kind (taking the definition used by Abramowitz \& Stegun \cite{abramowitz1965}). One thus obtains
\begin{equation}
  \label{eq:finalresultphioflambda}
  \phi(\lambda) = 
  \cases{
    0 \quad& \text{$|\lambda|\leq\lambda_\cc$ \,,}
\\
    \Lambda^{-1}(|\lambda|) \quad& \text{$|\lambda|\geq\lambda_\cc$ \,.}
  }
\end{equation}
For $|\lambda|<\lambda_\cc$, the scaled CGF $\phi(\lambda)$ and its associated optimal profiles are flat: one needs to consider large enough deviations of the current in order to actually observe a travelling trajectory. 
In order to check the existence of the transition and the form of the scaled CGF, 
we evaluated  $\phi(\lambda)$ from the maximal eigenvalue of the deformed operator~(\ref{eq:Wlambda}) of a lattice version of the dynamics (at small temperature and for a large number of sites).
Results are in good agreement with the present weak-noise approach (see Fig.~\ref{fig:phioflambda_comparision-WKB-diag}).

For the expansion close to the transition points, one finds for $\lambda=\lambda_\cc+\delta\lambda$ with $\delta\lambda>0$
\begin{equation}
  \label{eq:devphilambda-example}
 \phi(\lambda_\cc+\delta\lambda) = \frac{\pi \, \delta\lambda}{|\ln \delta \lambda|} + o\left( \frac{\delta\lambda}{|\ln \delta \lambda|} \right) \,.
\end{equation}
This leads for the average velocity $\overline{v}$ of the particle (or, equivalently, the average current),
\begin{equation}\label{eq:devphilambda-example-velocity}
\overline{v}(\lambda_\cc+\delta\lambda) = -\phi'(\lambda_\cc+\delta\lambda)
= -\frac{\pi}{|\ln \delta \lambda|} + o\left( \frac{1}{|\ln \delta \lambda|}\right) \,.
\end{equation}
As a result, the dynamical phase transition at $\lambda_\cc$ is formally continuous since $\overline{v}(\lambda) \rightarrow 0$ when $\lambda \to \lambda_\cc$.
However, for all practical purposes, the transition appears discontinuous as the convergence to zero is extremely slow.
This result is to be contrasted with the standard depinning transition of a particle in a `tilted' potential (\emph{i.e.}, a particle subjected to a conservative force plus a uniform non-conservative driving force). For a regular potential, the depinning transition is continuous with a critical exponent $1/2$~\cite{brazovskii_pinning_2004}.
The fact that the transition observed in the $\lambda$-biased dynamics is of a different nature shows that biasing the dynamics with $\lambda$ does not only add a non-conservative uniform driving force to the original dynamics, but rather modifies the original dynamics in a more complex way.
We describe in the next section how the $\lambda$-biased dynamics can be mapped onto a non-trivial effective driven process, which will allow us to better understand why the standard behaviour of the depinning transition of a particle in a potential is not recovered.


\section{Effective nonequilibrium dynamics of the conditioned equilibrium system}
\label{sec:effective-dyn}

As we discussed in paragraph~\ref{sec:large-devi-princ_large-time}, the biased dynamics is governed by the deformed Fokker--Planck operator $\WW_\lambda$ defined in~(\ref{eq:Wlambda}), which does not preserve probability. This observation is at the basis of population dynamics algorithms~\cite{grassberger_go_2002,giardina_direct_2006,tailleur_probing_2007} that allow one to study rare trajectories and to evaluate numerically the CGF by representing the probability loss or gain through selection rules between copies of the system, in the spirit of Quantum Monte Carlo algorithms~\cite{anderson_randomwalk_1975,makrini_diffusion_2007} (see \emph{e.g.}~\cite{giardina_simulating_2011} for a review).

In fact, as shown recently in~\cite{simon_construction_2009,popkov_asep_2010}  and in~\cite{jack_large_2010} (inspired by~\cite{evans_rules_2004,baule_invariant_2008}),
there exists a change of basis, based on the explicit knowledge of the left eigenvector of $\WW_\lambda$, that allows one to render the dynamics described by $\WW_\lambda$  \emph{probability-preserving}.
This defines an `auxiliary' or `effective' dynamics $\WW_\lambda^\eff$ which is asymptotically equivalent to the biased dynamics described by $\WW_\lambda$, as studied in great depth in Refs.~\cite{chetrite_nonequilibrium_2013,chetrite_nonequilibrium_2015,jack_effective_2015}. From a mathematical point of view, it is based on Doob's $h$-transform~\cite{miller_convexity_1961}.
The interest of this effective dynamics is that it provides a physical (probability-preserving) dynamics whose typical trajectories are equivalent to the rare trajectories of the original dynamics~(\ref{eq:Langevinepsilon}). Such effective dynamics can be defined for Langevin processes or for jump processes.
Explicit examples of such dynamics have been determined in exclusion processes~\cite{simon_construction_2009,popkov_asep_2010,popkov_transition_2011,belitsky_microscopic_2013}, in zero-range processes~\cite{harris_dynamics_2013,hirschberg_density_2015,chleboun_lower_2017}, in the current large deviation of Langevin dynamics~\cite{tsobgni_nyawo_large_2016} or in open quantum systems~\cite{carollo17a}.
They illustrate in general that the effective forces governing the dynamics described by $\WW_\lambda^\eff$ modify the original dynamics~(\ref{eq:Langevinepsilon}) on a global scale.

In this section, we recall how to identify the effective process as a Langevin dynamics with a force $F_\lambda^{\eff}(x)$ that defines a $\lambda$-modified probability-preserving dynamics~\cite{chetrite_nonequilibrium_2013,chetrite_nonequilibrium_2015}. We then show that the determination of $F_\lambda^{\eff}(x)$ can be done in a rather explicit way  in the weak-noise limit, without having to determine explicitly the left eigenvector.
We also explain how the determination of the effective process allows one to show that the small-noise and large-time limits can be exchanged in periodic systems for our LDF problem.

\subsection{Derivation of the effective force}
\label{sec:deriv-effect-force}

One defines $| L\rangle$ as the left eigenvector of $\WW _\lambda$ associated to the maximal eigenvalue $\varphi_\epsilon(\lambda)$.
Following Refs.~\cite{simon_construction_2009,popkov_asep_2010,jack_large_2010}, one introduces a diagonal operator $\hat L$ whose elements are the components of $| L\rangle$. Then, the definition $\langle L|\WW =\varphi_\epsilon(\lambda)\langle L|$ of the left eigenvector implies that
\begin{equation}
  \label{eq:defWstar}
  \langle -| \WW _\lambda^\eff = 0
\quad
\text{with}
\quad
  \WW _\lambda^\eff = \hat L \WW _\lambda \hat L^{-1} - \varphi_\epsilon(\lambda) \mathds{1} \,.
\end{equation}
One reads from this relation that the operator $\WW _\lambda^\eff$, which is equivalent to $\WW _\lambda$ up to a shift and a change of basis, is probability-preserving.
It thus provides a \emph{bona fide} dynamics whose properties are equivalent to the biased dynamics of the operator $\WW _\lambda$ (see paragraph~\ref{sec:interpretation_eff} for details on this equivalence).

\medskip

Let us now fix $\epsilon>0$ (not necessarily small) and send $\tf$ to infinity before taking the weak-noise limit.
The Perron--Frobenius theorem ensures that the components of the eigenvectors associated to the large eigenvalue can be taken to be strictly positive, 
which allows one to introduce a function $\tilde U(x)$ such that 
the left eigenvector reads $L(x)=\ee^{-\frac 1\epsilon \widetilde U(x)}$.
Note that this relation is simply a definition of $\tilde U(x)$, which may depend on $\epsilon$ at this stage. However, one expects\footnote{%
This can be formally shown for instance by expanding $\tilde U(x)$ in a power series as $\tilde U(x)=\tilde U_0(x)+\epsilon\tilde U_1(x)+\ldots$ following the standard WKB procedure \cite{Olver1974}.
}
that in the weak-noise limit
$\epsilon \to 0$, $\tilde U(x)$ becomes independent of $\epsilon$.%

One finds by direct computation that
\\
\begin{minipage}{1.0\linewidth}%
\begin{align}
  \WW _\lambda^\eff P (x)= 
&
\frac{1}{2} \epsilon  P''(x)
+  \big[ -F(x)+\lambda  g(x)+\tilde{U}'(x)\big] P'(x)
\nonumber
\\
&
+
\frac{1}{2\epsilon}
\Big[
\lambda ^2 g(x)^2-2 \lambda  h(x) -2 \epsilon \big( \varphi_\epsilon (\lambda )+  F'(x) - \lambda  g'(x)\big)
\nonumber
\\
&
\phantom{ +\frac{1}{2\epsilon}\Big\{ }\,
-2 F(x) \tilde{U}'(x)+\tilde{U}'(x)^2+2 \lambda  g(x) (\tilde{U}'(x)-F(x))+\epsilon  \tilde{U}''(x) \Big] P(x) \,.
\label{eq:WefflambdaPphi}
\end{align}%
\end{minipage}%
\\
Using the eigenvector equation for $L(x)$
\begin{equation}
\frac{1}{2} \epsilon  L''(x)+\big[F(x)-\lambda  g(x)\big] L'(x)
+ \frac{\lambda}{\epsilon} \left[ \frac{1}{2} \lambda  g(x)^2 -h(x)-F(x) g(x) \right]  L(x)
=
\varphi_\epsilon(\lambda)  L(x)
\label{eq:leftev-eq}
\end{equation}
which can be rewritten in terms of $\tilde U(x)$ as
\begin{equation}
-\frac{1}{2} \tilde{U}''(x)
+
\frac{1}{\epsilon}
  \left[ \frac{1}{2} \big(\lambda  g(x)+\tilde{U}'(x)\big) \big(\lambda  g(x)+\tilde{U}'(x)-2 F(x)\big)-\lambda  h(x) \right]
=
\varphi_\epsilon(\lambda) \,,
\label{eq:eqUtilde_phi}
\end{equation}
one eliminates $\varphi_\epsilon(\lambda)$ in~(\ref{eq:WefflambdaPphi}) and one finds that $\WW _\lambda^\eff$ indeed takes the form of a probability-preserving Fokker--Planck evolution operator~\cite{chetrite_nonequilibrium_2013,chetrite_nonequilibrium_2015}:
\begin{equation}
 \WW _\lambda^\eff \cdot
  =-\partial_x\Big[\big(F(x)-\lambda  g(x)-\tilde{U}'(x)\big) \cdot\Big]
  +\frac{1}{2}\epsilon\partial^2_x\cdot\,\,.
\label{eq:Wefflambda-final}
\end{equation}
It describes the evolution of a particle subjected to a force $F^\eff(x)=F(x)-\lambda  g(x)-\tilde{U}'(x)$.
We note that the contribution $h(x)$ to the additive observable $A$ defined in~\eqref{eq:obs} does not appear explicitly in~(\ref{eq:Wefflambda-final}) but is still present implicitly through the potential $\tilde U(x)$ defined from the left eigenvector $L(x)$.

\subsection{Effective dynamics in the weak-noise limit}

Noting that $\varphi_\epsilon(\lambda)  \sim \frac 1 \epsilon \phi(\lambda)$
in the weak-noise limit $\epsilon \to 0$, and assuming that $\tilde{U}$ becomes independent of $\epsilon$ in this limit (as is usually the case in this WKB procedure \cite{Olver1974}),
the differential equation~(\ref{eq:eqUtilde_phi}) for $\tilde U(x)$ becomes an ordinary, quadratic equation for $\tilde U'(x)$,
\begin{equation}
\frac{1}{2} \big(\lambda  g(x)+\tilde{U}'(x)\big) \big(\lambda  g(x)+\tilde{U}'(x)-2 F(x)\big)-\lambda  h(x) =
\phi(\lambda) \,,
\label{eq:eqUtilde_phi2}
\end{equation}
whose solution reads
\begin{equation}
  \label{eq:solUprimeWKB}
  \tilde U'(x) = F(x)-\lambda  g(x) -\sigma \sqrt{F(x)^2+2 \lambda  h(x)+2 \phi (\lambda )} \,,
\end{equation}
where $\sigma = \pm 1$ is an unknown sign that will be determined later on.
Hence, the knowledge of $\phi(\lambda)$ allows for the determination of $\tilde U'(x)$, provided one is able to select the correct sign in Eq.~(\ref{eq:solUprimeWKB}). This can be done by evaluating the effective force $F^\eff(x)$.
Inserting Eq.~(\ref{eq:solUprimeWKB}) in the generic expression~(\ref{eq:Wefflambda-final}) of the effective Fokker--Planck operator, one finds
\begin{equation}
 \WW _\lambda^\eff \cdot
  =
  -\partial_x\Big[\sigma \sqrt{F(x)^2+2 \lambda  h(x)+2 \phi (\lambda )}\: \cdot\Big]
  +\frac{1}{2}\epsilon\partial^2_x\cdot\,\,.
\label{eq:Wefflambda-weak-noise}
\end{equation}
It corresponds to the evolution of a particle subjected to an effective force
\begin{equation}
F^\eff(x)=\sigma \sqrt{F(x)^2+2 \lambda  h(x)+2 \phi (\lambda )} \,.
\label{eq:FeffWKB}
\end{equation}
The two possible signs correspond to the two possible cases of paragraphs~\ref{sec:propagative-trajs} and~\ref{sec:case-purely-current} when the optimal trajectory is either increasing or decreasing in time. 
{We will see below that $\sigma$ is given by $\sigma=-{\rm sign}(B)$, consistently with the results of paragraph~\ref{sec:propagative-trajs}.}

We have thus shown that in the weak-noise asymptotics, the explicit knowledge of the complete left eigenvector $|L\rangle$ is not required in order to determine the effective force $F^\eff(x)$: one only needs to know the scaled CGF $\phi(\lambda)$.
Interestingly, for periodic systems, Eq.~(\ref{eq:solUprimeWKB}) also provides a way to determine $\phi(\lambda)$, without using the optimisation procedure described in Sec.~\ref{sec:case-peri-syst}.
In a periodic system, $\tilde{U}(x)$ is a periodic function of period $1$, so that $\int_0^1 \tilde{U}'(x)\,dx=0$. From Eq.~(\ref{eq:solUprimeWKB}), we thus have
\begin{equation} \label{eq:integral:Uprime}
\int_0^1 dx\,\Big( F(x)-\lambda  g(x) - \sigma \sqrt{F(x)^2+2 \lambda  h(x)+2 \phi (\lambda )} \Big) =0
\end{equation}
and one recovers Eq.~(\ref{eq:determin:Estar}), given the definitions~(\ref{eq:def:B:K}) and~(\ref{eq:devpotV}) of the parameter $B$, the function $R$ and the potential $\mathcal{V}(x)$, as well as the identification of $\phi(\lambda)$ with $\mathcal{E}_{\sigma}^{\ast}$ when Eq.~(\ref{eq:determin:Estar}) has a solution
---~see Eqs.~(\ref{eq:phi:lambda:prop:E}) and~(\ref{eq:phi:phiprop}).
Following the same reasoning as the one that leads to Eq.~(\ref{eq:sigma:signB}), we recover that $\sigma=-{\rm sign}(B)$.

Note that recovering the same result as in Sec.~\ref{sec:case-peri-syst} is non-trivial, because here we have made no optimisation of the action at finite time $\tf$, but rather taken the infinite-time limit from the outset, by using first a spectral analysis (which yielded the eigenvector equation~\eqref{eq:leftev-eq}) and then a weak-noise expansion to go from~(\ref{eq:eqUtilde_phi}) to~(\ref{eq:eqUtilde_phi2}). In other words, we have exchanged the order of  the large-time and the weak-noise limit. It is interesting to see, as we previously mentioned, that both orders of limits yield the same result. 
Let us emphasise that this result strongly relies on the Perron--Frobenius theorem, which states that the eigenvector $|L\rangle$ associated to the largest eigenvalue of $\mathbb W_\lambda$ only has strictly positive components (up to a sign convention) so that it can be written in an exponential form $L(x)=\ee^{-\frac 1\epsilon \widetilde U(x)}$ (with real $\widetilde U(x)$), while other eigenvectors do not have all components of the same sign, and can thus not be written in such an exponential form. Looking for an eigenvector in an exponential form thus automatically selects the eigenvector associated to the largest eigenvalue thanks to the Perron--Frobenius theorem, without any explicit optimisation procedure.

\begin{figure}[t]
\begin{center}
  \includegraphics[width=.55\columnwidth]{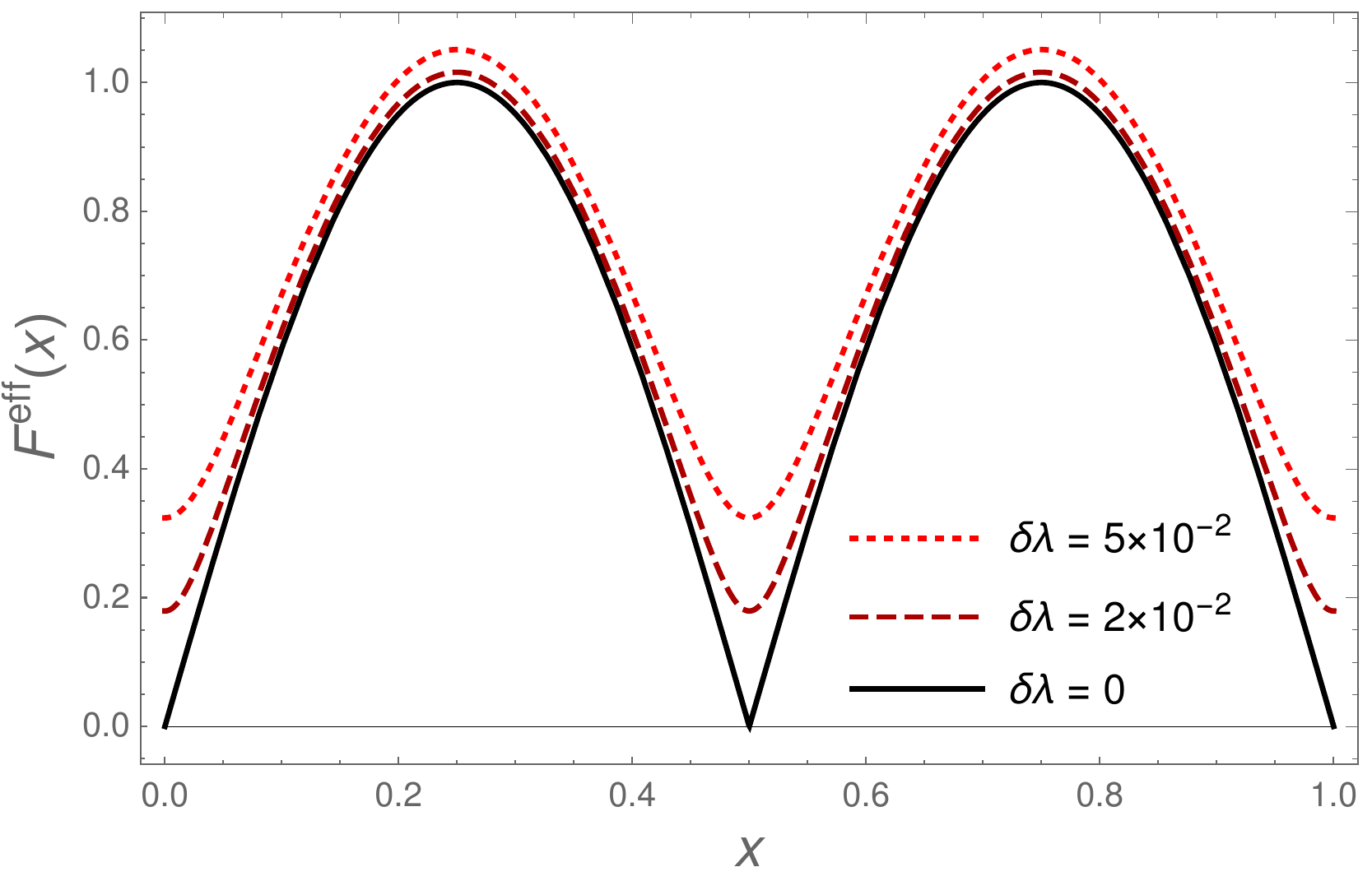}\\
  \includegraphics[width=.48\columnwidth]{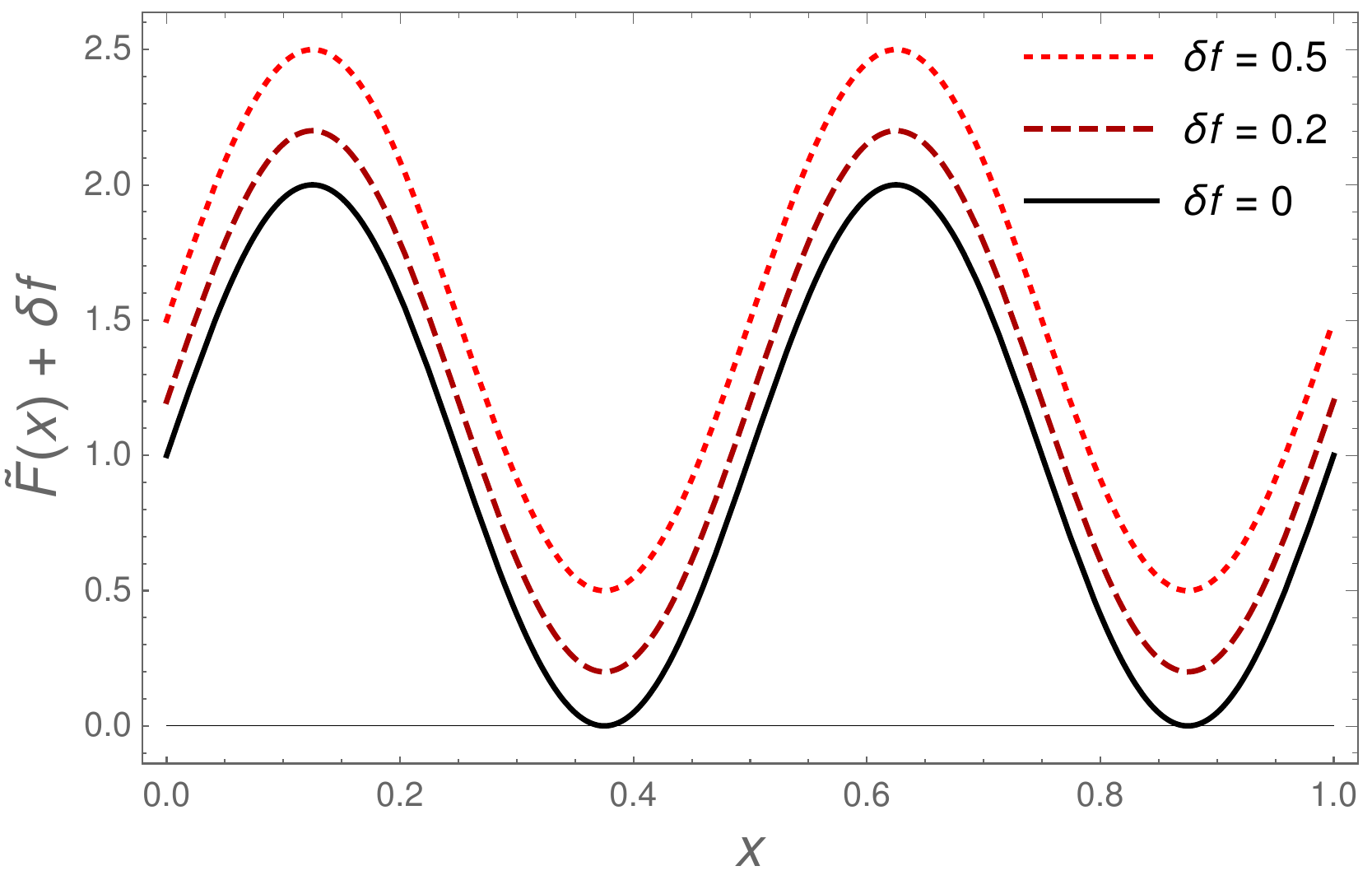}\hfill
  \includegraphics[width=.48\columnwidth]{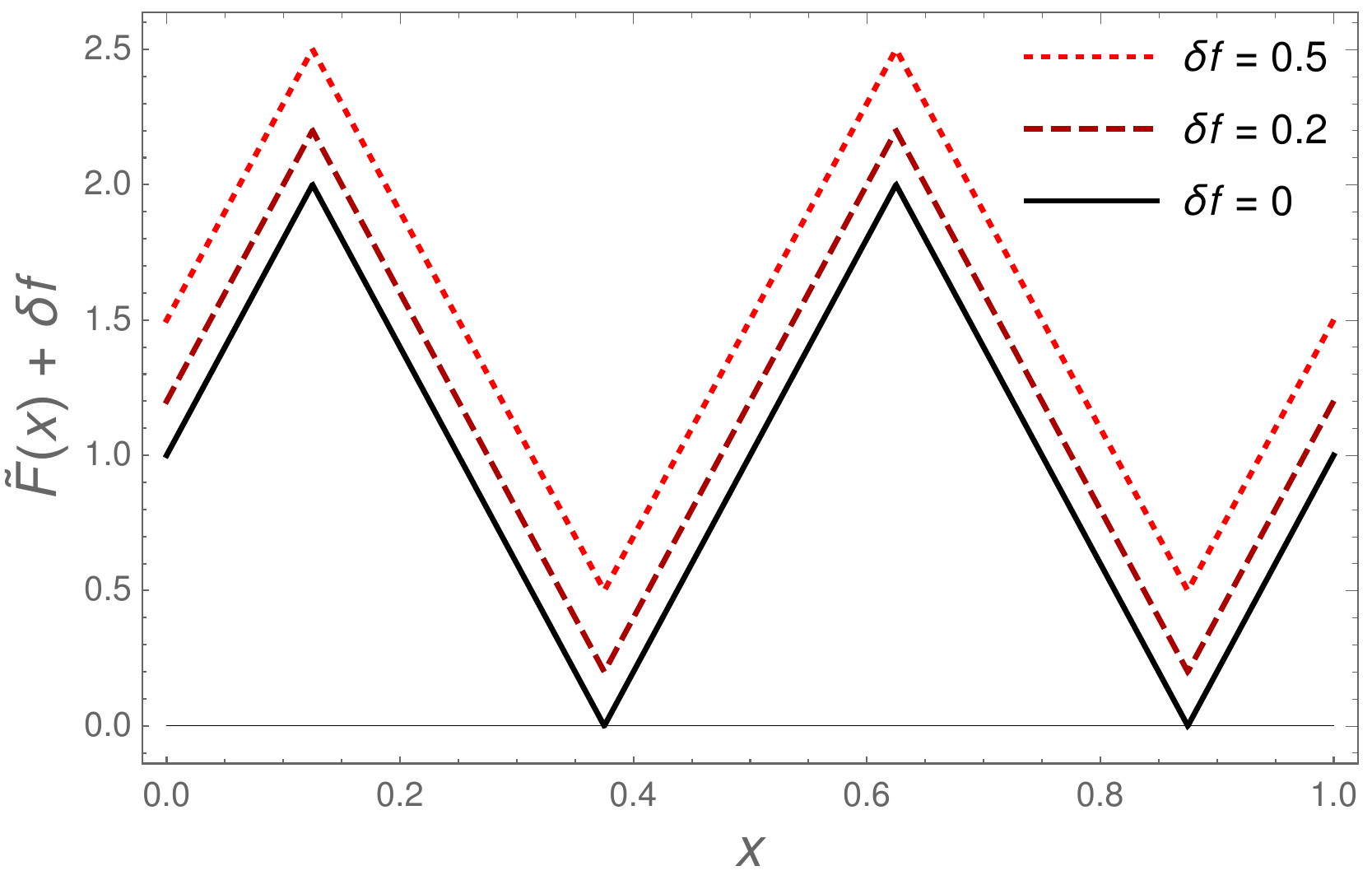}
\end{center}
\caption{%
  Comparison, on an example, of the criticality of the dynamical phase transition and of standard depinning transitions in 0d. (\textbf{Top}) The effective force  $F^\eff(x)$ at   $\lambda=\lambda_\cc+\delta\lambda$ deforms and becomes a cuspy function of $x $ close to the stationary points that develop as $\delta\lambda\to0^+$, for the example model studied at the end of Sec.~\ref{sec:propagative-trajs}.
(\textbf{Bottom left}) In the standard depinning transition in 0d, the depinning transition occurs when a regular force $\tilde F(x)$ possesses no stationary point any more when driven by a uniform force $\delta f>0$. In this case, the force $\tilde F(x)+\delta f$ is a regular function of $x$ and this implies that the transition is second-order~\cite{brazovskii_pinning_2004}, in contrast to our dynamical phase transition of  interest.
(\textbf{Bottom right}) If instead the force presents a cusp at all values of $\delta f\geq 0$ close to the $\delta f=0$ stationary point, the transition becomes first-order~\cite{brazovskii_pinning_2004}, which is also different from our dynamical phase transition.
\label{fig:compar-DPT-depinning}
}
\end{figure}

For a conservative force $F(x)$ and current-type additive observables (\emph{i.e.}, $h(x)=0$), the condition $\sigma=-{\rm sign}(B)$ boils down (if $g(x)>0$) to $\sigma=-{\rm sign}(\lambda)$, leading to an effective force
\begin{align} \label{eq:Feff:curr:obs}
F^\eff(x) &= -{\rm sign}(\lambda) \, \sqrt{F(x)^2 + 2 \phi (\lambda )}
\qquad &(|\lambda| > \lambda_\cc) \,,\\
F^\eff(x) &= -{\rm sign}(\lambda) \, |F(x)|  &(|\lambda| \le \lambda_\cc) \,.
\end{align}
Note that $F^\eff(x)$ can be decomposed into a uniform non-conservative part
\begin{equation} \label{eq:Feff:nonconserv}
f_\eff = \int_0^1 F^\eff(x) \, dx
= \int_0^1 F(x) \, dx - \lambda \int_0^1 g(x) \, dx
\end{equation}
(where the last equality results from Eq.~(\ref{eq:integral:Uprime}))
and a space-dependent conservative part
\begin{equation} \label{eq:Feff:conserv}
-U_\eff' = F^\eff(x) - f_\eff \,.
\end{equation}
Note that the integrals in Eq.~(\ref{eq:Feff:nonconserv}) should be understood as spatial averages
(we recall that the length of the system is chosen as $L=1$).
In the specific case when the observable $A$ is the current (\emph{i.e.}, $g(x)=1$ and $h(x)=0$) and $F(x)$ is a conservative force, one recovers that $f_\eff =-\lambda$. Yet, the conservative part
$-U_\eff'$ does not reduce to the original force $F(x)$.
This can be seen explicitly by computing perturbatively the effective force $F^\eff(x)$ in the large $\lambda$ limit, yielding
\begin{equation}
 F^\eff(x) \stackrel[\lambda\rightarrow\infty]{}{=} -\lambda
- \frac{1}{2\lambda} \left( F(x)^2 - \int_0^1 F(x')^2 \, dx' \right) + o \left( \frac{1}{\lambda}\right)
\end{equation}
From this last equation, the decomposition of the effective force $F^\eff(x)$ into a uniform non-conservative force $f_\eff =-\lambda$ and a conservative force becomes
\begin{equation}
-U_\eff' = - \frac{1}{2\lambda} \left( F(x)^2 - \int_0^1 F(x')^2 \, dx' \right) + o \left( \frac{1}{\lambda}\right) .
\end{equation}
The associated periodic potential $U_\eff$ reads
\begin{equation}
U_\eff(x) = \frac{1}{2\lambda} \left( \int_0^x F(x')^2 \, dx' 
- x \int_0^1 F(x')^2 \, dx' \right) + o \left( \frac{1}{\lambda}\right) .
\end{equation}
On the other side, it is instructive to determine how the effective force $F^\eff(x)$ is modified as the dynamical transition is approached: 
we illustrate in Fig.~\ref{fig:compar-DPT-depinning} how a cusp singularity appears in $F^\eff(x)$ as $\delta\lambda\to0^+$ for $\lambda=\lambda_\cc+\delta\lambda$ in the example system studied at the end of Sec.~\ref{sec:propagative-trajs}. One sees from Eq.~(\ref{eq:Feff:curr:obs}) that $F^\eff(x)$ is a regular function of $x$ as long as $\delta\lambda>0$ but develops a cusp singularity at its stationary points as $\delta\lambda\to0^+$, explaining why the transition is not of the same nature as that of the standard depinning transition~\cite{brazovskii_pinning_2004}.

To understand on a more general ground the relation between such depinning transition and the dynamical phase transition, we now consider the more generic case of an observable $A$ with arbitrary $g$, $h\equiv 0$ and a force $F(x)$ presenting a stationary point $x_0$. We assume that $F$ can be expanded around $x_0$ as $F(x)=(x-x_0)F_0+o(x-x_0)$.
In the effective dynamics, optimal trajectories are subjected to the effective force defined in Eq.~(\ref{eq:FeffWKB}) which reads as follows close to the stationary point\footnote{We assume here that $\sigma=+1$ without loss of generality.}:
\begin{equation}
 F^\eff(x)\simeq \sqrt{[(x-x_0)F_0]^2+2\phi(\lambda)}.
 \label{eq:Feffexpandx0}
\end{equation}
As the dynamical phase transition is approached ($\phi(\lambda)\to0$ as $\lambda\to\lambda_\cc^+$), this implies that the trajectories of the effective dynamics
spend a longer and longer time close to $x_0$, meaning that the dynamics is mainly governed by the approximate form~\eqref{eq:Feffexpandx0} of the effective force.

Effective trajectories are governed by the equation  $\dot{x}(t) =  F^\eff(x(t))$, whose solution reads
\begin{equation}
 x(t)=x_0\pm\sqrt{2\phi(\lambda)}\: \frac{\sinh (F_0 t)}{F_0},
\end{equation}
(up to an arbitrary translation in time) in the regime where the approximation~(\ref{eq:Feffexpandx0}) holds.
As usual for the depinning transition in 0d problems~\cite{brazovskii_pinning_2004}, the average velocity of the trajectory (counted positively) is $|\bar v|\sim L/T$, with $L$ the spatial period ($L=1$ in our settings) and $T$ the period in time.
Close to the transition, the time period $T$ is for instance estimated from
\begin{equation}
  \label{eq:eqT}
  x(T/2)-x(-T/2) = L = 1
\end{equation}
(since $x(0)=x_0$ with our choice of the origin of time, so that $x(t)-x_0$ is an odd function of time).
In the large-time limit, one finds from~(\ref{eq:eqT}) that at dominant order in $\phi(\lambda)\to 0$, one has
$T\sim\frac{1}{F_0}|\ln\phi(\lambda)|$. 
Consequently, the average velocity of the trajectory behaves as:
\begin{equation}
 |\bar v|\sim\frac{F_0}{|{\ln{\phi(\lambda)}}|},
 \label{eq:vbarlambda}
\end{equation}
in good agreement with the result~(\ref{eq:devphilambda-example-velocity}) for the example of the $\sin(2\pi x)$ force. In all, we have shown that the behaviour~(\ref{eq:Feffexpandx0}) of the effective force close to the stationary point of the effective depinning problem governs the logarithmic form of the velocity close to the transition for an arbitrary current-type additive observable~$A$. 

Furthermore, if the additive observable $A$ is the velocity of the particle, the relation $v(\lambda) = -\phi'(\lambda)$ together with~\eqref{eq:vbarlambda} leads to:
\begin{equation}
 \phi'(\lambda)\sim\frac{1}{|\ln{\phi(\lambda)}|}
\qquad
\textnormal{ for $\lambda\to\lambda_c^+$}\:.
\end{equation}
Close the transition point, writing $\lambda= \lambda_c+\delta\lambda$, one can integrate this equation and get:
\begin{equation}
 \phi(\lambda)\ln{\delta\lambda}-\phi(\lambda)=-\delta\lambda.
\end{equation}
so that at dominant order for $\delta\lambda\to0^+$
\begin{equation}
 \phi(\lambda)\sim\frac{\delta\lambda}{|\ln{\delta\lambda}|}.
\end{equation}
which is compatible with the result~(\ref{eq:devphilambda-example}) obtained for the example model $F(x)=\sin(2\pi x)$.

In conclusion, the effective depinning transition problem fully describes the behaviour of the expansion of $\phi(\lambda)$ near the transition~(\ref{eq:devphilambda-example}), with an effective force of the form~(\ref{eq:Feffexpandx0}).

\subsection{{Interpretation of the effective process in the path-integral formulation}}

It is now interesting to come back to the path integral formulation introduced in Sec.~\ref{sec:rare-traj-cond} to discuss the relationship between the effective nonequilibrium process and the original process biased by $\lambda$.
To simplify the discussion, we specialise here to a conservative force
$F(x)=-U'(x)$, so that we compare the effective nonequilibrium process with a $\lambda$-biased equilibrium process. The generalisation to a non-conservative force field $F(x)$ is straightforward.

A Langevin process with the effective nonequilibrium force $F^\eff(x)$,
\begin{equation}
\label{eq:Langevinepsilon:eff}
 \dot{x}(t)=F^\eff(x(t))+\sqrt{\epsilon}\,\eta(t)\,\,,
\end{equation}
leads to an Onsager--Machlup action which, in the small-noise limit, takes the form
\begin{equation}
S_\eff[x(t),\tf] = \int_0^{\tf} \frac{1}{2} \big(\dot{x}-F^\eff(x) \big)^2 \;dt
\:.
\end{equation}
Expanding the square in the action and using the expression~(\ref{eq:FeffWKB}) of the effective force $F^\eff(x)$, we end up with
\begin{equation} \label{eq:def:Seff}
S_\eff[x(t),\tf] = \int_0^{\tf} \left\{ \frac{1}{2} \dot{x}^2 + \frac{1}{2} F(x)^2
+\lambda h(x) + \phi(\lambda) - \dot{x} F^\eff(x) \right\} \, dt \,.
\end{equation}
Then, Eqs.~(\ref{eq:FeffWKB}) and~(\ref{eq:integral:Uprime}) imply that
\begin{equation}\label{eq:efforce:to:orforce}
\int_0^{\tf} \dot{x} F^\eff(x) \, dt = - \int_0^{\tf} \lambda \dot{x} g(x)\, dt
\end{equation}
so that the action of the nonequilibrium process reads
\begin{equation}
S_\eff[x(t),\tf] = \int_0^{\tf} \left\{ \frac{1}{2} \dot{x}^2 + \frac{1}{2} F(x)^2
+\lambda \big( h(x) + \dot{x} g(x) \big) + \phi(\lambda) \right\} \, dt \,.
\end{equation}
Hence,
\begin{equation} \label{eq:compar:Seff:Sla}
S_\eff[x(t),\tf] = \Sla[x(t),\tf] + \phi(\lambda) \tf
\end{equation}
and the action $S_\eff[x(t),\tf]$ of the effective nonequilibrium process identifies, up to a constant which is irrelevant to the statistics of $x(t)$, with the action $\Sla[x(t),\tf]$ of the biased process given in Eq.~(\ref{eq:S-lambda1}).
This confirms that the effective nonequilibrium process defined by the force field $F^\eff(x)$ describes the same statistics of trajectories as the original dynamics biased by $\lambda$.

Let us emphasise that in Eq.~(\ref{eq:compar:Seff:Sla}), the actions $S_\eff[x(t),\tf]$ and $\Sla[x(t),\tf]$ compare the nonequilibrium dynamics characterised by a force $F^\eff(x)=-U_\eff'-\lambda$, with a $\lambda$-biased dynamics in the original conservative force field $F(x)=-U'(x)$.
Since the non-conservative force $f_\eff =-\lambda$ generates a current, one may also wonder how the driven dynamics defined by $F^\eff(x)$ compares with the $\lambda$-biased dynamics in the \emph{same} potential $U_\eff(x)$ (which, we recall, differs from the original potential $U(x)$).
In other words, does the driven dynamics sample uniformly all the rare equilibrium trajectories exhibiting the same value of the average current?
The action $\Sla^\eff[x(t),\tf]$ of the $\lambda$-biased dynamics in the potential $U_\eff(x)$ reads
\begin{equation}
\Sla^\eff[x(t),\tf] = \int_0^{\tf} \left\{ \frac{1}{2} \Big( \dot{x}+U_\eff'(x) \Big)^2 + \lambda \dot{x} \right\} \, dt
\end{equation}
while the action $S_\eff[x(t),\tf]$ of the driven process defined in Eq.~(\ref{eq:def:Seff}) can be rewritten as
\begin{equation}
S_\eff[x(t),\tf] = \int_0^{\tf} \left\{ \frac{1}{2} \Big( \dot{x} + U_\eff'(x) \Big)^2 + \lambda \dot{x} + \lambda U_\eff'(x) + \lambda^2 \right\} \, dt \,.
\end{equation}
The difference $S_\eff[x(t),\tf]-\Sla^\eff[x(t),\tf]$ then reads
\begin{equation}
S_\eff[x(t),\tf]-\Sla^\eff[x(t),\tf] = \int_0^{\tf}
\Big( \lambda U_\eff'(x) + \lambda^2 \Big) \, dt \,.
\end{equation}
Trajectories in the driven process are thus not sampled uniformly among the rare equilibrium configurations with the same current, but are reweighted with an exponential prefactor
\begin{equation}
\exp\left\{ -\frac{\lambda}{\epsilon} \int_0^{\tf}  U_\eff'(x) \, dt \right\}
\end{equation}
(the $\lambda^2$ term has been discarded since it is independent of $x$).
Only in the trivial case when the conservative force vanishes do the two statistics coincide.

\subsection{{Equivalence between the effective driven process and the $\lambda$-biased process}}
\label{sec:interpretation_eff}

{We have shown above, using the path-integral formalism in the specific case of a conservative force field $F(x)$, that the effective driven process described by $\WW_\lambda^\eff$ is equivalent to the original $\lambda$-biased process described by $\WW_\lambda$. We provide here for completeness a more general and formal proof of this equivalence in an operatorial formalism, following Refs.~\cite{jack_large_2010,giardina_direct_2006,nemoto_population-dynamics_2016}. The force field $F(x)$ is here no longer assumed to be conservative.}

We start by defining a `$\lambda$-ensemble' as a normalised average 
$
\langle \mathcal \,\cdot\,\rangle_\lambda^{[0,\tf]}$
in the biased dynamics, namely
\begin{equation}
  \label{eq:deflambdaensemble}
  \langle \mathcal O[x(t)] \rangle_\lambda^{[0,\tf]} = 
  \frac
  {\big\langle \mathcal O[x(t)]\, \ee^{-\frac\lambda\epsilon A(\tf)}\big\rangle}
  {\big\langle  \ee^{-\frac\lambda\epsilon A(\tf)}\big\rangle}
  \:,
\end{equation}
where the observable $\mathcal O$ depends on the trajectory. We made explicit the time interval on the l.h.s.~because the statistical properties of the $\lambda$-ensemble at times close to $\tf$ are different from those in the ``bulk'' of the time interval $[0,\tf]$.

Let us now focus on an observable $\mathcal O[x(t)]=\mathcal O(x(t_1))$ which depends only on the position of the particle at a time $t_1\in[0,\tf]$.
Denoting by $\hat \mathcal O$ the diagonal operator whose components are the values of $\mathcal O(x)$, one has by definition
\begin{equation}
  \label{eq:operatorformlambdaens}
  \big\langle\mathcal O(x(t_1))\big\rangle_\lambda^{[0,\tf]} = 
  \frac
  {\langle -  | \ee^{(\tf-t_1)\WW_\lambda} \: \hat \mathcal O\: \ee^{t_1\WW_\lambda} | P_\ii \rangle}
  {\langle -  | \ee^{\tf\WW_\lambda}| P_\ii \rangle}
  \:.
\end{equation}
Then, if both $t_1$ and $\tf-t_1$ are large compared to the inverse of the gap of $\WW_\lambda$, that is to say, if $t_1$ is in the bulk of the time interval $[0,\tf]$, one can use the asymptotic behaviour~(\ref{eq:large-t-behaviour_operator}) of $\ee^{t \WW_{\lambda}}$, leading to
\begin{equation}
  \label{eq:operatorformlambdaens_bulk}
  \big\langle \mathcal O(x(t_1)) \big\rangle_\lambda^{[0,\tf]} 
  \ \stackrel[\tf\to\infty]{}{\longrightarrow} \
  {\langle L  | \hat \mathcal O  | R \rangle}
  =
  \int \mathcal O(x) L(x) R(x)\:dx
  \:.
\end{equation}
In other words, as well known~\cite{giardina_direct_2006,nemoto_population-dynamics_2016}, the intermediate-times $\lambda$-ensemble statistics is governed by the product of the left- and right-eigenvectors of $\WW_\lambda$.

Consider now the effective dynamics described by the operator $\WW_\lambda^\eff$. From its definition~(\ref{eq:defWstar}), one sees that the left- and right-eigenvectors associated to its largest eigenvalue $0$ are respectively equal to $|-\rangle$ and $|LR\rangle$.
In analogy with~(\ref{eq:large-t-behaviour_operator}), the large-time behaviour of the propagator $\ee^{\tf\, \WW_\lambda^\eff}$ is thus given by
\begin{equation}
  \label{eq:large-t-behaviour_operator_Weff}
  \ee^{\tf\,\WW_\lambda^\eff} 
  \ \stackrel[\tf\to\infty]{}{\sim} \
  |LR\rangle\langle -| \,,
\end{equation}
where $|LR\rangle \equiv \hat{L}|R\rangle$.
Hence, similarly to~(\ref{eq:operatorformlambdaens_bulk}), one finds that the average
$
  \langle\,\cdot\,\rangle_\eff^{[0,\tf]}
$
of an observable for the effective dynamics is given in the steady state by
\begin{equation}
  \label{eq:operatorformeffectivedyn}
  \big\langle\mathcal O(x(t_1))\big\rangle_\eff^{[0,\tf]} 
  \ \stackrel[\tf\to\infty]{}{\longrightarrow} \
  {\langle -  | \hat \mathcal O  | LR \rangle}
  =
  \int \mathcal O(x) L(x) R(x)\:dx \,,
\end{equation}
which is equal to the corresponding $\lambda$-ensemble average~(\ref{eq:operatorformlambdaens_bulk}).
The statistical properties of the biased dynamics, described by~(\ref{eq:deflambdaensemble}), are thus equal to those of the effective dynamics at any time $t_1$ in the bulk of $[0,\tf]$.

Using the identity 
$
\ee^{t\,\WW_\lambda^\eff}
=
\ee^{-t\,\varphi_\epsilon(\lambda)}\:
\hat L \,
\ee^{t\,\WW_\lambda}
\hat L^{-1}
$
inferred from the definition~(\ref{eq:defWstar}) of the effective operator,
the previous reasoning can be readily extended to observables of the form $\mathcal O[x(t)]=\mathcal O(x(t_1),x(t_2),\ldots)$ depending on the position of the particle at different times $t_1$, $t_2$, $\ldots$ which are all in the bulk of the interval $[0,\tf]$ (but which can be arbitrarily close to each other).
This corresponds to the notion of trajectorial equivalence of the biased and the effective measure developed by Chetrite and Touchette~\cite{chetrite_nonequilibrium_2013,chetrite_nonequilibrium_2015}.


\section{Conclusion and outlook}
\label{sec:outlook}

In this work we have identified an effective probability-conserving dynamics turning the rare trajectories of a stochastic process into the typical histories of an explicit modified dynamics, in the case of a particle diffusing in a periodic one-dimensional generic force under a weak thermal noise. In this way, by using large-deviation techniques, we have determined the form of the force that a particle effectively withstands when conditioned to bear an atypical fluctuation for a long duration.
{Interestingly, the resulting effective nonequilibrium process does not only differ from the original $\lambda$-biased dynamics by the addition of a ($\lambda$-dependent) uniform driving force, but the conservative part of the force is also ``renormalised'' by the presence of $\lambda$, even in the simple case when the observable $A(\tf)$ is the average current.
Considering this result from a reversed perspective, we have also argued that a particle in a potential driven by a uniform non-conservative force does not sample in a uniform way the set of all trajectories having the same current ---~perhaps at odds with naive expectations.

Along the way,} we have analysed the fluctuations of time-integrated current-type observables in a periodic system. These display a rich phenomenology associated to the existence of dynamical phase transitions between a static fluctuating phase, characterised by a flat sCGF, and a phase with time-periodic travelling trajectories, associated to a sCGF being equal to the energy of a natural optimisation problem ---~which takes the form of a conservative Hamiltonian dynamics.
Furthermore, we have described an alternative way to compute the sCGF without using the variational techniques derived from the weak-noise analysis of the path-integral representation,
that allowed us to show how the large-time and the weak-noise limits commute.

The obtained results also open a direction of research to characterise fluctuations in a given system by engineering a new system subjected to an additional external force. Such an approach has been used in recent studies on adaptive algorithms, based for instance on a feedback procedure to evaluate the effective force~\cite{nemoto_population-dynamics_2016,nemoto_finite-size_2017,ray_exact_2017,ferre_adaptive_2018}, improving the computational efficiency. The weak-noise regime has been seldom studied (it is in fact know to present specific difficulties~\cite{nemoto_population-dynamics_2016}), and the results we present in this paper could help to understand large-time fluctuations and their associated phenomenology both in experiments and simulations.

\section*{Acknowledgments}
Financial supports from Spanish projects FIS2017-84256-P (Ministerio de Economia y Competitividad) and FPU13/05633 as well as from the grant IDEX-IRS `PHEMIN' of the Université Grenoble Alpes are acknowledged.
VL~acknowledges support by the ERC Starting Grant No.~680275 MALIG and the ANR-15-CE40-0020-03 Grant LSD. 

\section*{References}

\addcontentsline{toc}{section}{References}
\bibliographystyle{plain_url}

\bibliography{effective-dynamics_small-noise-langevin}

\end{document}